\documentclass[preprint,pra,aps,nofootinbib,longbibliography]{revtex4-1}
\usepackage{graphicx}

\begin{document}

\title{Visualizing quantum entanglement in Bose-Einstein condensates without state vectors}
\author{Russell B. Thompson}
\email{thompson@uwaterloo.ca}
\affiliation{Department of Physics \& Astronomy and Waterloo Institute for Nanotechnology, University of Waterloo, 200 University Avenue West, Waterloo, Ontario, Canada N2L 3G1}
\date{\today}

\begin{abstract}
Ring polymer self-consistent field theory is used to calculate the critical temperatures and heat capacities of an ideal Bose gas for an order of magnitude more particles than previously reported. A $\lambda$-transition indicative of Bose-Einstein condensation is observed as expected. Using a known proof of spatial mode entanglement in Bose-Einstein condensates, a relationship between boson exchange and quantum entanglement is established. This is done without the use of state vectors, since ring polymer quantum theory uses instead a thermal degree of freedom, sometimes called the ``imaginary time'', to map classical statistical mechanics onto non-relativistic quantum mechanics through the theorems of density functional theory. It is shown that quantum phenomena, such as Bose-Einstein condensation, boson exchange, entanglement and contextuality, can be visualized in terms of merging and separating ring polymer threads in thermal-space. A possible extension to fermions is mentioned.
\end{abstract}

\maketitle

\section{Introduction}
Quantum state vectors are the fundamental mathematical quantities on which non-relativistic quantum mechanics is based. Postulated to contain all the information that is knowable about a quantum system \cite{Bransden2000, CohenTannoudji1977, Shankar1994}, state vectors suffer from practical and philosophical difficulties such as the collapse of the wave function and instantaneous correlations between distant  isolated parts of a system arising from quantum entanglement. The collapse of the wave function, often called the measurement problem, occurs when the unitary deterministic evolution of the state vector as described by the Schr\"{o}dinger equation is abruptly changed to the Born rule upon measurement. If one assumes that the wave function describes an individual system (rather than an ensemble), then this is unphysical. Some possible solutions to the measurement problem include  spontaneous collapse theories \cite{Bassi2003}, the many-worlds interpretation \cite{Vaidman2022} and pilot wave theory \cite{Goldstein2010}. Correlations from entanglement seem similarly unphysical, in that they include causal correlations between distant non-interacting systems that are instantaneous \cite{Weinberg2014, *Weinberg2014pre}. Steven Weinberg (among many others) has summarized these difficulties, and has proposed that the density matrix, rather than the state vector, be used as the fundamental description of physical reality \cite{Weinberg2014, *Weinberg2014pre}. He defined the density matrix to be more general than just an ensemble of state vectors with a range of probabilities, pointing out that the density matrix contains much less information than the underlying state vectors. Weinberg was not the first to contemplate state-free quantum mechanics; Feynman and Hibbs pondered using the path integral method to avoid the mention of energy levels, and suggested a motivation for doing this would be to achieve a deeper understanding of physical processes \cite{Feynman1965}. 

Feynman's path integral method of quantum mechanics underpins ring polymer quantum theory, which describes quantum particles as classical ring trajectories embedded in a fictitious thermal-space, avoiding energy levels and quantum states \cite{Shankar1994, Ceperley1995, Feynman1953b, Roy1999b}. The fictitious thermal dimension is often referred to as the ``imaginary time'' because if one Wick rotates the time variable $it/\hbar$ into the inverse temperature $1/k_BT$ in the space-time path integral expression for the probability amplitude for the dynamics of a quantum particle, one gets the equilibrium partition function for the quantum particle. This holds only if one constrains all trajectories in the partition function to begin and end at the same spatial point. Projected into three dimensions, these trajectories form ``rings'', and the partition function is the same as the \emph{classical} partition function of ring polymers. Ring polymer quantum theory has grown into a small industry of methods \cite{Ceperley1995, Roy1999b, Zeng2014, Althorpe2009, Habershon2013}, including self-consistent field theory (SCFT). SCFT gives a propagator, which is essentially a position-basis density matrix, as the fundamental quantity in ring polymer quantum theory \cite{Thompson2019, Thompson2020, Sillaste2022, Thompson2022, Thompson2023, LeMaitre2023a, LeMaitre2023b, Kealey2024}. This answers a question posed by Weinberg \cite{Weinberg2014}: If the density matrix is not merely an ensemble of state vectors, then what is it? SCFT gives an ontology for quantum particles and the density matrix in terms of classical thermal loops. Specifically, the SCFT approach models quantum particles as classical thermal threads that have well-defined ring conformations at every instant of time, and solves for equilibrium ensemble averages of these thermal trajectories using classical statistical mechanics. More than this, SCFT has been shown to be equivalent to quantum density functional theory (DFT) \cite{Thompson2019}, and the theorems of DFT \cite{Hohenberg1964} guarantee equivalence between classical SCFT predictions and those of static non-relativistic quantum mechanics. This quantum-classical isomorphism \cite{Chandler1981} can be extended to include time and temperature dependent phenomena through further DFT theorems \cite{Mermin1965, Runge1984}. This means that there is a rigorous basis for using classical ring polymers in thermal-space-time as a model for quantum particles. This is done without state vectors and so is free of the measurement problem, as will be discussed in section \ref{sec-Discussion}.

This classical ontology for non-relativistic quantum mechanics has been tested through numerical atomic and molecular calculations \cite{Thompson2019, Thompson2020, Sillaste2022, LeMaitre2023a, LeMaitre2023b, Kealey2024}. It gives explanations for quantum phenomena including the measurement problem, quantum kinetic energy, the uncertainty principle, the stability of atoms, molecular bonding, tunnelling, the double slit experiment and geometric phase including the Aharonov-Bohm effect \cite{Sillaste2022, Thompson2023}. In the high temperature limit, the SCFT picture reduces to classical DFT with the extra thermal dimension spontaneously obeying a cylinder condition \cite{Thompson2019}. Thus, with no new assumptions, classical ring polymer SCFT meets the criteria for Kaluza theory \cite{Kaluza1921, Wesson1997, Dinov2022} through which electromagnetism emerges from a five dimensional (5D) derivation of general relativity \cite{Thompson2022}.

For fermions, quantum exchange in three spatial dimensions (3D) has been shown to be equivalent to classical excluded volume in four thermal-space dimensions (4D), including topologically forbidden configurations of the thermal trajectory \cite{Thompson2020, Kealey2024}. Thus the Pauli exclusion principle naturally emerges in SCFT and gives rise to shell structure in atomic calculations \cite{LeMaitre2023a, LeMaitre2023b}. For bosons, closely related ring polymer simulation techniques have been applied to many systems \cite{Ceperley1995, Roy1999b, Zeng2014, Althorpe2009, Habershon2013}. Polymer field-theoretic simulation (FTS), which is closely related to SCFT, has been applied to the ideal boson gas \cite{Fredrickson2022}, interacting bosons \cite{Fredrickson2020, Fredrickson2022}, thermodynamic engines with boson working fluids \cite{Fredrickson2023b}, spin microemulsions of coupled Bose-Einstein condensates \cite{Fredrickson2023c} and non-equilibrium boson systems \cite{Fredrickson2023}. To date however, SCFT itself has not been applied to bosons.

In this work, the specific heat for the ideal boson gas is calculated using SCFT. An ideal gas of bosons is the simplest possible quantum system with boson exchange, and SCFT offers advantages over other methods for the study of this system. Previous work using FTS has, to date, not calculated the heat capacity or related quantities of the ideal boson gas both above and below the critical point of a Bose-Einstein condensate (BEC) \cite{Fredrickson2022}. Also, when working in the canonical ensemble, SCFT results are exact for finite particle numbers and agree with previous formulas and numerical results \cite{Mullin2000, Mullin2003, Vakarchuk2001, Matsubara1951, Feynman1972}, whereas standard grand canonical methods fail for low temperatures and/or small numbers of particles \cite{DelMaestro2020, Kocharovsky2010, Kocharovsky2014, Kocharovsky2015}. Using SCFT, it is simple to calculate the heat capacity above, below and through the critical point of the ideal gas condensate with an order of magnitude more particles than has previously been reported. A BEC is normally defined as a state of matter in which a large number of bosons all occupy the same ground state. Since the SCFT formalism does not use wave functions, it is shown in this contribution that a BEC can instead be understood in terms of the merging and separation of ring polymer loops in classical 4D thermal-space. This connects boson exchange directly with entanglement since every BEC is in a highly entangled state \cite{Simon2002, Heaney2009, Laflorencie2016, Schmied2016}. This exchange-entanglement connection, coupled with classical ring polymer visualization, provides a means to picture entanglement in the BEC system. The main results of this paper are therefore that a classical ring polymer model for quantum particles gives that same results as standard QM, and that quantum entanglement in the ideal Bose gas can be shown to be equivalent to the mechanism of boson exchange given by Feynman.

\section{Theory}  \label{sec-Theory}
The ring polymer approach to quantum mechanics was introduced by Feynman \cite{Feynman1953b} and is now a standard approach in quantum theory \cite{Ceperley1995, Roy1999b, Zeng2014, Althorpe2009, Habershon2013}. From the textbook of Shankar \cite{Shankar1994}, the partition function for a single quantum particle of mass $m$ at a temperature $T$ in a potential $U({\bf r})$ is
\begin{equation}
Z(\beta) = \int d{\bf r} \int_{\bf r}^{\bf r} \mathcal{D} {\bf r} \exp \left\{ -\int_0^{\beta} \left[\frac{m}{2\hbar^2} \left(\frac{d{\bf r}}{ds}\right)^2 + U({\bf r}(s)) \right] ds    \right\}  \label{part1}
\end{equation}
where $\beta = 1/k_BT$, $k_B$ is Boltzmann's constant, and $\hbar$ is Planck's reduced constant. In equation (\ref{part1}), the position of the particle ${\bf r}$ is parametrized in $s$, and the path integral explores all trajectories that begin and end at the position ${\bf r}$ in the 4D space of position and inverse thermal energy. If $s$ is considered to be embedded in 3D rather than being an independent thermal dimension, then equation (\ref{part1}) is identical to the configurational integral of the standard coarse-grained model for a classical ring polymer \cite{Matsen2006, Fredrickson2006}. Thus one can choose to view quantum particles as classical threads in a 4D thermal-space. The thermal dimension is often referred to as the imaginary time because if one replaces $\beta$ with $it/\hbar$ in equation (\ref{part1}) (a Wick rotation), one gets the probability amplitude for the trajectory of a quantum particle in space-time \cite{Feynman1953b}. It can be seen that the thermal degree of freedom has the properties of a dimension in the sense that it appears as mathematically equivalent to time.

In the canonical ensemble, the Helmholtz free energy of a single classical ring polymer (quantum particle) system can be found from (\ref{part1}) and phrased in a field-theoretic way using SCFT. In brief, Hubbard-Stratonovich transformations are used in SCFT to convert polymer partition functions from particle-based to field-based descriptions; full details can be found in references \cite{Matsen2006, Matsen2002, Fredrickson2006, Qiu2006, Schmid1998}. Ignoring quantum exchange effects for the moment, the free energy of the partition function (\ref{part1}) generalized to $N$ interacting particles is
\begin{equation}
F = -\frac{N}{\beta} \ln Q_1 + U[n] - \int d{\bf r} w({\bf r})n({\bf r}) + \frac{1}{\beta} \ln N!  . \label{FE1}
\end{equation}
The first term contains the single particle partition function $Q_1$, which is subject to a field $w({\bf r})$ determined by interactions with all other particles and external potentials. The second term $U[n]$ describes all the potentials which give rise to $w({\bf r})$, and is a functional of the spatially inhomogeneous quantum particle number density, $n({\bf r})$. The field $w({\bf r})$ thus controls the density distribution of the quantum particles. The third term subtracts off the energy of the field $w({\bf r})$, and the last term is a constant for fixed $N$ and is usually dropped in SCFT.

The free energy (\ref{FE1}) is used as an action to be extremized through functional differentiation with respect to the functions $w({\bf r})$ and $n({\bf r})$ generating a set of self-consistent equations:
\begin{eqnarray}
n({\bf r},\beta) &=& \frac{N}{Q_{1}(\beta)} q({\bf r},{\bf r},\beta)  \label{dens1}   \\
Q_{1}(\beta) &=& \int q({\bf r},{\bf r},\beta) d{\bf r}    \label{Q1}  \\
w({\bf r}) &=& \frac{\delta U[n]}{\delta n({\bf r})}  \label{w1}  \\
\frac{\partial q({\bf r}_0,{\bf r},s)}{\partial s} &=& \frac{\hbar^2}{2m} \nabla^2 q({\bf r}_0,{\bf r},s) - w({\bf r},\beta) q({\bf r}_0,{\bf r},s)  . \label{diff1}
\end{eqnarray}
The main governing equation of ring polymer quantum SCFT is equation (\ref{diff1}), which is a modified diffusion equation for a real and non-negative propagator $q({\bf r}_0,{\bf r},s)$ that obeys the initial condition 
\begin{equation}
q({\bf r}_0,{\bf r},0) = \delta({\bf r}-{\bf r}_0)   .  \label{init1}
\end{equation}
This propagator needs to be solved for the values of the diagonal, $ q({\bf r},{\bf r},s)$, to find the particle density $n({\bf r})$ in equation (\ref{dens1}). 

\begin{figure}
\centering
\begin{tabular}{cc}
\includegraphics[scale=0.2]{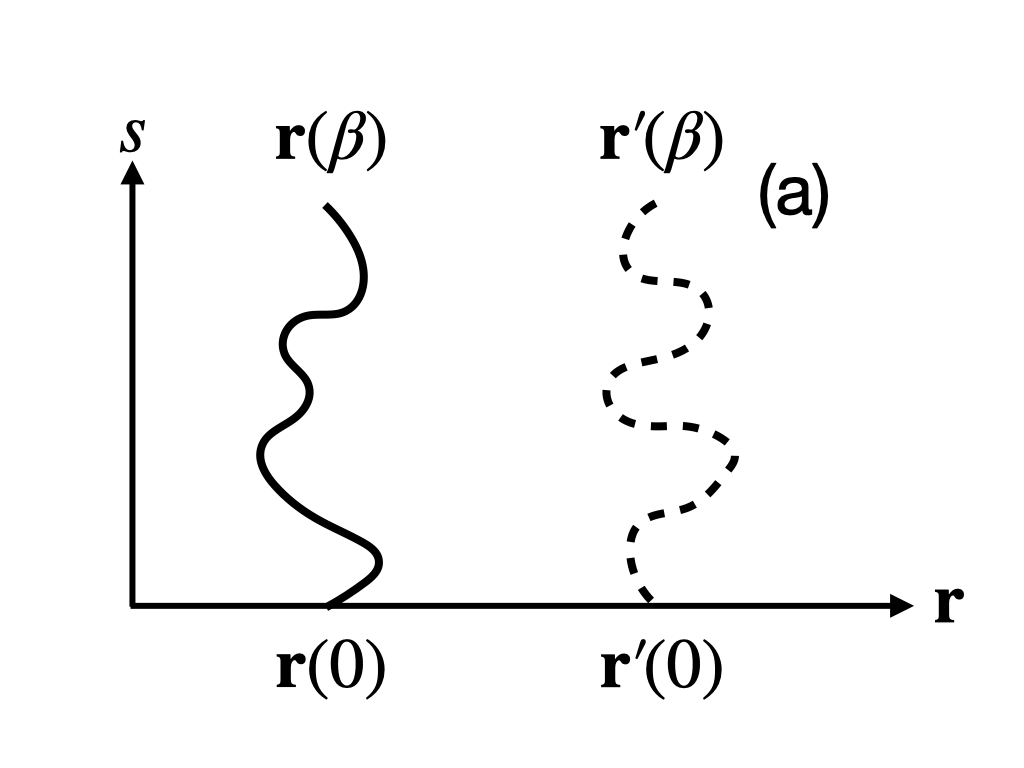} &
 \includegraphics[scale=0.2]{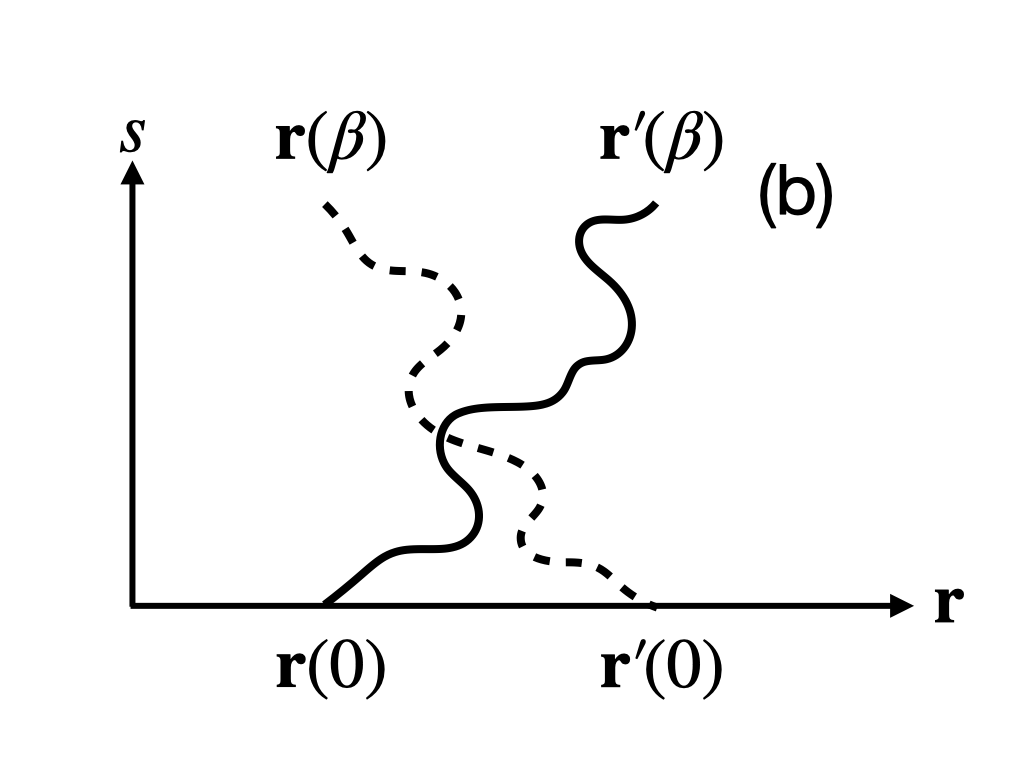} \\
\includegraphics[scale=0.2]{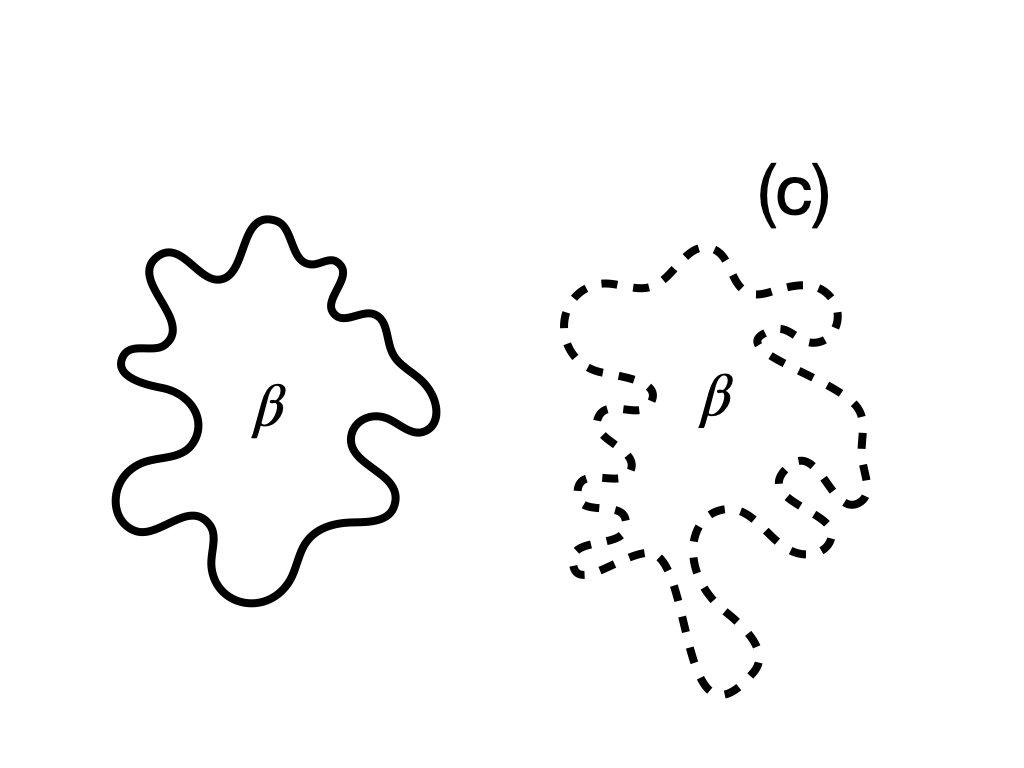} &
\includegraphics[scale=0.2]{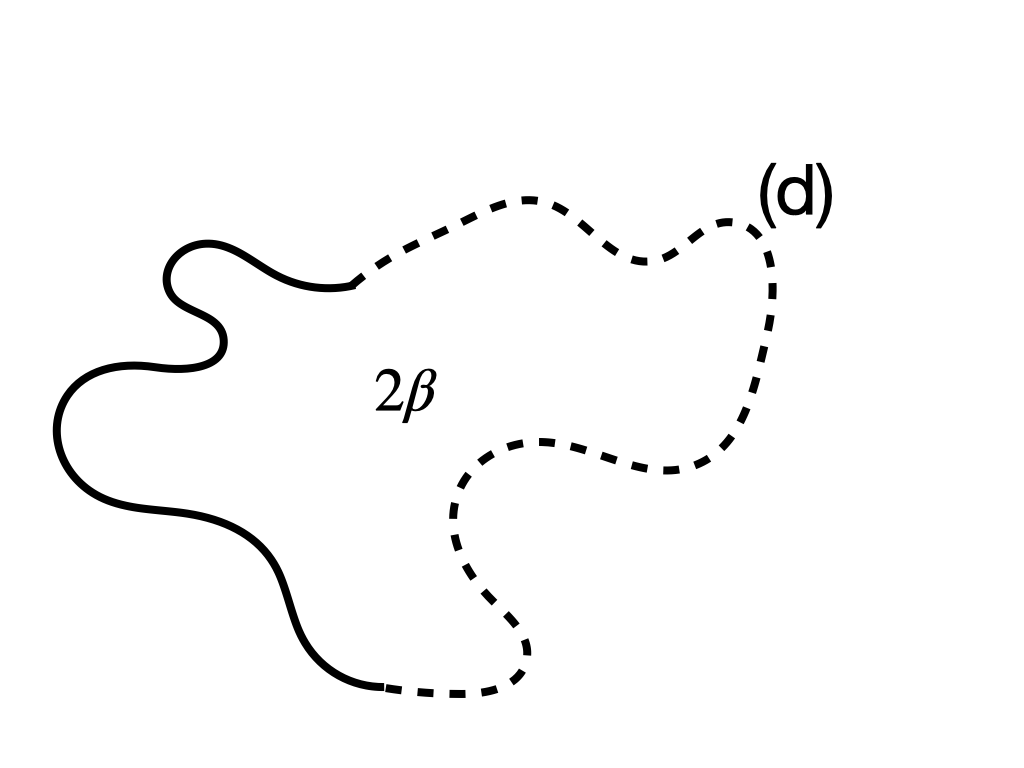} 
\end{tabular}
\caption{Schematics of imaginary time trajectories for two quantum ``ring polymers''. (a) At imaginary time $s=0$, one trajectory (solid curve) starts at position ${\bf r}$ and the other (dashed curve) starts at ${\bf r}^\prime$. At $s=\beta$, both curves return to their initial positions;  ${\bf r}$ for the solid trajectory and ${\bf r}^\prime$ for the dashed trajectory. Three dimensional space is collapsed onto the horizontal position axis, and the vertical axis is the inverse thermal energy. (b) Boson exchange corresponds to the solid and dashed curves switching final positions at $s=\beta$: the solid curve starts at  ${\bf r}$ and ends at ${\bf r}^\prime$ and the dashed curve starts at ${\bf r}^\prime$ and ends at ${\bf r}$. (c) Projection of the solid and dashed trajectories onto real space without imaginary time, showing that they each appear as closed rings of contour length $\beta$ in ${\rm I\!R}^3$. (d) Projection of the solid and dashed trajectories with exchanged end points, showing that they appear as a single ring polymer of contour length $2\beta$ in ${\rm I\!R}^3$.}
\label{fig-exchange}
\end{figure}

Up to this point, the classical ring polymer equations are identical to those of a system of quantum particles without exchange. Including boson exchange corresponds to allowing thermal trajectories to exchange final positions, as illustrated in figure \ref{fig-exchange} panels (a) and (b) for a two particle situation.
As derived and discussed in references \cite{Thompson2023, Kealey2024, Fredrickson2023, Feynman1953b}, this quantum exchange is equivalent to a classical ring polymer system where rings can merge and separate, back and forth between larger and smaller rings as depicted in figure \ref{fig-exchange} panels (c) and (d). The single particle partition function (\ref{Q1}) should therefore be replaced by a multi-particle partition function to account for merging and separating microstates, as derived for the two-particle case in reference \cite{Thompson2023}. The formulas for two, three and four particle partition functions are
\begin{eqnarray}
Q_{2}(\beta) &=& Q_{1}(\beta)^2 + Q_{1}(2\beta) \label{Q2}  \\
Q_{3}(\beta) &=& Q_{1}(\beta)^3 + 3Q_{1}(\beta)Q_{1}(2\beta) + 2Q_{1}(3\beta)  \label{Q3} \\
Q_{4}(\beta) &=& Q_{1}(\beta)^4 + 6Q_{1}(\beta)^2Q_{1}(2\beta) + 3Q_{1}(2\beta)^2 + 8Q_{1}(\beta)Q_{1}(3\beta) +  6Q_{1}(4\beta)  \label{Q4} .  
\end{eqnarray}
It is important to notice that all orders of partition functions $Q_N$ are expressible as functions of the single particle partition function $Q_1$. More details on the origins of equations (\ref{Q2})-(\ref{Q4}) can be found in the appendix.

For an ideal gas of bosons, there are no interactions apart from exchange, and the field given by equation (\ref{w1}) can be set to zero without loss of generality. The diffusion equation (\ref{diff1}) can then be solved analytically to give
\begin{equation}
q({\bf r}_0,{\bf r},\beta) = \left(\frac{m}{2\pi\hbar \beta}\right)^\frac{3}{2} \exp \left[ \frac{-m({\bf r} - {\bf r}_0)^2}{2\hbar^2\beta}\right]   . \label{diff2}
\end{equation}
For ring polymers, the initial and final positions must be the same, so ${\bf r} = {\bf r}_0$ and $q({\bf r}_0,{\bf r},\beta)$ becomes constant in space
\begin{equation}
q(\beta) =  \left(\frac{m}{2\pi \hbar^2 \beta}\right)^\frac{3}{2}  = \frac{1}{\Lambda^3}    \label{diff3}
\end{equation}
where $\Lambda$ is the thermal wavelength. The single particle partition function, equation (\ref{Q1}), becomes just 
\begin{equation}
Q_{1}(\beta,V) = \frac{V}{\Lambda^3}   \label{Q1b}
\end{equation}
where $V$ is the volume of the system in the canonical ensemble. This almost trivial diffusion equation solution is the main result of this contribution in that it shows that classical SCFT will give identical results as purely quantum methods for all the equilibrium properties of a boson ideal gas. To see this, note that the equilibrium partition function for a finite system of $N$ particles (ring polymers) will be
\begin{equation}
Z_{N} = \frac{Q_{N}(\beta)}{N!}   \label{Z1}
\end{equation}
where the pattern of equations (\ref{Q2})-(\ref{Q4}) can be continued to give the general expression for $Q_{N}(\beta)$ derived in the appendix:
\begin{equation}
Q_{N}(\beta) = \sum_{j=1}^{P(N)} c_j Q_{1}(\beta)^{l_j}  \label{QN}
\end{equation}
with 
\begin{equation}
c_j = \frac{N!}{\prod_{k \in \delta_j} k^{\frac{5}{2} g(k)} g(k)!}  . \label{cj}
\end{equation} 
Equations (\ref{QN}) and (\ref{cj}) are defined in terms of quantities taken from the theory of integer partitions. The integer partition of a positive integer $N$ is the number of ways of writing $N$ as a sum of positive integers, denoted by $P(N)$. For example, the integer $N=4$ can be written as 1+1+1+1, 2+1+1, 2+2, 3+1 or 4, so $P(4) = 5$. This makes sense in the current context since the rings polymers can merge and separate into larger or smaller rings as described by the integer partition. Each of the integer partition sums can be written as a set, for example 2+1+1 can be written as $\{2,1\}$ with degeneracies $g(2) = 1$ and $g(1) = 2$, that is, 2 appears once and 1 appears twice. In equation (\ref{QN}), the sum is over all the sets of the integer partition $P(N)$ with $l_j$ being the cardinality of the $j$th set, accounting also for degeneracies. Continuing the example of $N=4$, if the set $\{2,1\}$ is considered the second set of the integer partition, labelled as $j=2$, then $\delta_2 = \{2,1\}$ with $l_2 = 3$ because there are three integers in the sum 2+1+1. In equation (\ref{cj}) for the coefficient $c_j$, the product in the denominator is over all elements of the $j$th set. The appendix walks through the construction of equations (\ref{QN}) and (\ref{cj}) from the point of view of classical ring polymers. Equations (\ref{QN}) and (\ref{cj}) are equivalent to expressions given by Mullin \cite{Mullin2000} and earlier by Matsubara \cite{Matsubara1951} and Feynman \cite{Feynman1972}. More details on the integer partition in the context of the statistical mechanics of various quantum gases is given by Zhou and Dai \cite{Zhou2018}.

The heat capacity for a system of $N$ particles, $C_N$, can be found as a function of temperature or volume from the partition function expressions (\ref{Z1}), (\ref{QN}) and (\ref{cj}) using
\begin{equation}
C_N = k_B\beta^2 \frac{\partial^2 \ln Z_N}{\partial \beta^2}  .   \label{CN}
\end{equation}
It is more convenient to change the independent variable in equation (\ref{CN}) from $\beta$ to $Q_1$, the single particle partition function given by equation (\ref{Q1b}), as this scales temperature with the volume and therefore captures both the temperature and density dependence of the heat capacity. Writing equation (\ref{CN}) in terms of $Q_1$ and as a specific heat (heat capacity per particle) gives
\begin{equation}
\frac{C_N}{k_BN} = \frac{9}{4} \frac{Q_1^2}{Z_{N}} \left[\frac{\partial^2 Z_{N}}{\partial Q_1^2} - \frac{1}{Z_{N}}\left(\frac{\partial Z_{N}}{\partial Q_1}\right)^2 \right] + \frac{15}{4} \frac{Q_1}{Z_{N}} \frac{\partial Z_{N}}{\partial Q_1}  .   \label{CN2}
\end{equation}
The derivatives in equation (\ref{CN2}) can be solved analytically as sums using equation (\ref{QN}):
\begin{eqnarray}
\frac{\partial Q_N}{\partial Q_1} &=& \sum_{j=1}^{P(N)} c_j l_j Q_1^{l_j-1}  \label{dQN} \\
\frac{\partial^2 Q_N}{\partial Q_1^2} &=& \sum_{j=1}^{P(N)} c_j l_j (l_j-1) Q_1^{l_j-2} .  \label{d2QN} 
\end{eqnarray}
Given a number of particles $N$ and a value of $Q_1$, equations (\ref{Z1})-(\ref{cj}) and (\ref{CN2})-(\ref{d2QN}) give the specific heat.  

The Matsubara formula (\ref{Z1})-(\ref{cj}) is usually only practical for $N<10^2$ particles. Beyond that, the integer partition becomes numerically cumbersome. Mullin and Fern\'{a}ndez have pointed out that the Matsubara formula given by (\ref{QN}) and (\ref{cj}) is the solution of a well known recurrence relation first derived by Landsberg \cite{Mullin2003, Landsberg1961}. In the notation of this paper, the recurrence relation for the partition function $Z_N$ is
\begin{equation}
Z_N = \frac{Q_1}{N} \sum_{l=1}^N \frac{Z_{N-l}}{l^\frac{3}{2}}  \label{recurr}
\end{equation}
with derivatives
\begin{eqnarray}
Z_N^\prime &=& \frac{1}{N} \sum_{l=1}^N \frac{1}{l^\frac{3}{2}} \left( Q_1 Z_{N-l}^\prime +Z_{N-l} \right)  \label{dZ}  \\
Z_N^{\prime\prime} &=& \frac{1}{N} \sum_{l=1}^N \frac{1}{l^\frac{3}{2}} \left( Q_1 Z_{N-l}^{\prime\prime} +2 Z_{N-l}^\prime \right)  \label{d2Z}  
\end{eqnarray} 
where primes denote differentiation with respect to $Q_1$. The specific heat can be calculated using equation (\ref{CN2}) to a higher value of $N$ than is possible using the Matsubara equations (\ref{QN}) and (\ref{cj}) because there is no need to compute the integer partition. 

For  $N$ values much larger than $10^3$, $Z_N$ becomes very big and the recursion relation (\ref{recurr}) is numerically limited by floating point number size. Park and Kim have solved this problem by noting that one needs only the ratios of $Z_N$ and its derivatives in the expression for the heat capacity (\ref{CN2}) \cite{Park2010}. These ratios are not large, so Park and Kim rewrite the recursion relation for $Z_N$ as 
\begin{equation}
Z_N = \prod_{n=1}^N f_n  \label{recurr2}
\end{equation}
where
\begin{equation}
f_N = \sum_{n=0}^{N-1} \left[ \frac{Q_1/(N-n)^\frac{3}{2}}{N\prod_{j=n+1}^{N-1} f_j}\right]  .   \label{recurr3}
\end{equation}
The necessary ratios in equation (\ref{CN2}) are therefore
\begin{eqnarray}
\frac{Z_N^\prime}{Z_N} &=& \sum_{n=1}^N \frac{1}{n^\frac{5}{2} \prod_{k=N-n+1}^N f_k}    \label{rdZ}  \\
\frac{Z_N^{\prime\prime}}{Z_N} &=& \sum_{n=1}^N \frac{1}{n^\frac{5}{2}} \left(\sum_{m=1}^{N-n} \frac{1}{m^\frac{5}{2} \prod_{k=N-n-m+1}^N f_k}\right)  .   \label{rd2Z} 
\end{eqnarray} 
Specific heat calculations which spontaneously show BEC are presented in the next section using all three methods described above.

\section{Results}  \label{sec-Results}
\begin{figure}
\includegraphics[scale=0.25]{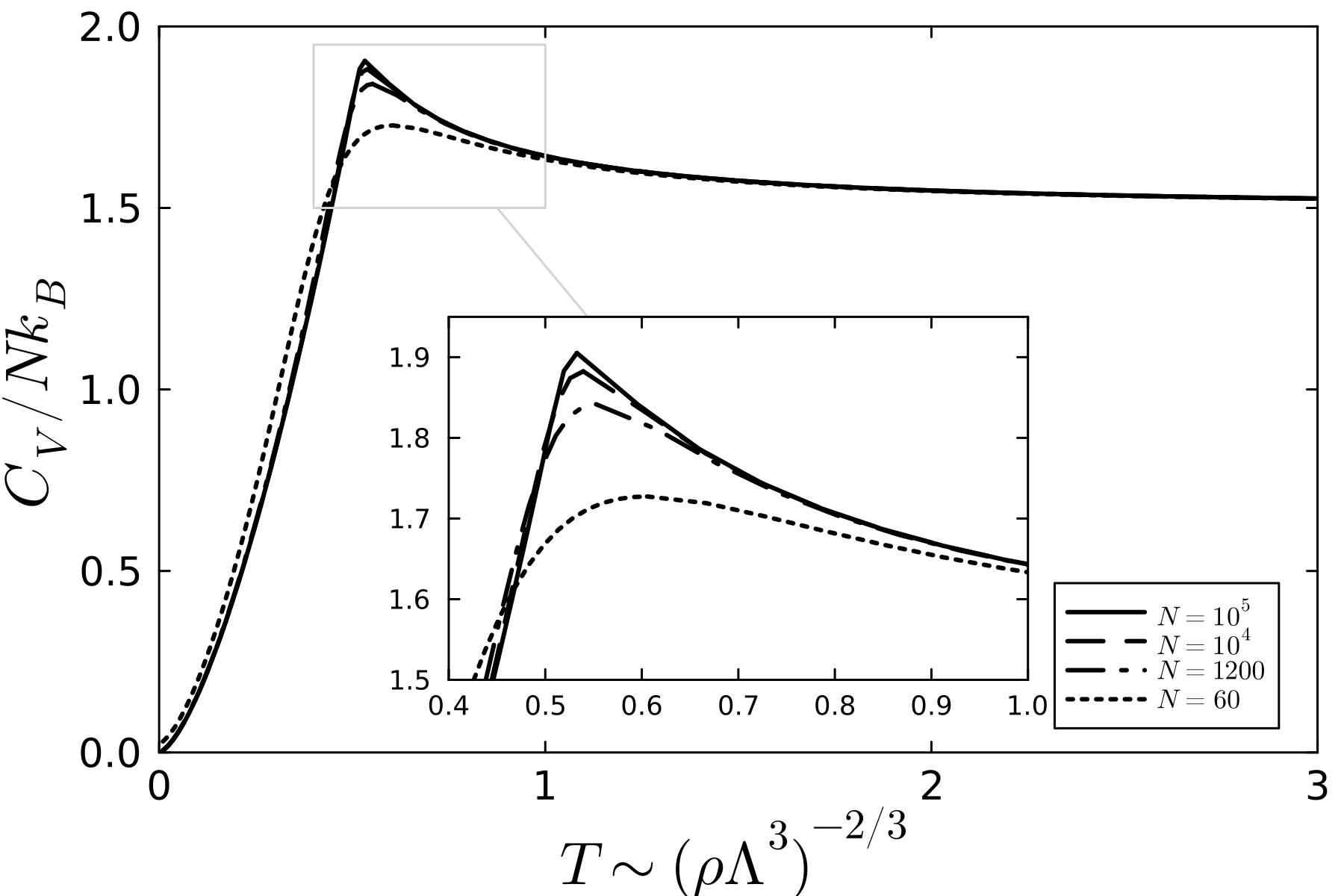} 
\caption{Plots of specific heat as a function of scaled temperature for a variety of particle numbers. The inset enlarges the vicinity of the critical point.}
\label{fig-heat}
\end{figure}

Plots of specific heat vs $Q_1$ for the ideal Bose gas are shown in figure \ref{fig-heat} for a range of $N$. All three numerical methods -- integer partition up to $N=60$, recursion up to $N=1200$ and ratio recursion up to $N=10^5$ --  produced exactly the same results for overlapping $N$ values.
Vakarchuk and Rovenchak have tabulated both heat capacities and critical temperatures for the ideal Bose gas up to $N=1200$ \cite{Vakarchuk2001}. SCFT results agree with Vakarchuk and Rovenchak for both quantities to all reported digits: Table \ref{tab:rholam} gives scaled critical temperature values (product of particle density $\rho = N/V$ and the cube of the critical thermal wavelength at the  maximum of the $C_N/N k_BT$ vs scaled temperature curve) and specific heats at the critical temperature for different particle numbers.
\begin{table}
\caption{Scaled critical temperature values (product of density and the cube of the critical thermal wavelength $\rho \Lambda_c^3$) and specific heat (heat capacity per particle) at the critical point for different particle numbers.}
\vspace{0.5cm}
\begin{tabular}{|c|c|c|}
\hline 
\ \ $N$ \ \ & \ \ \ \ $\rho \Lambda_c^3$ \ \ \ \ & \ \ $C_N^{\rm max}/k_B N$  \\
\hline \hline 
1 & --- & 1.500 \\
\hline 
10 & 1.717 & 1.624 \\
\hline
$10^2$ & 2.200 & 1.753  \\
\hline
$10^3$ & 2.423 & 1.837  \\
\hline
$10^4$ & 2.525 & 1.882  \\
\hline
$10^5$ & 2.572 & 1.905  \\
\hline
\end{tabular}
\label{tab:rholam}
\end{table}
In table \ref{tab:rholam}, the SCFT calculations have been extended up to $N = 10^5$, which is an order of magnitude higher than given by Park and Kim \cite{Park2010}\footnote{Park and Kim report some thermodynamic properties up to $N=10^6$ particles, but only give heat capacity data up to $N=10^4$.} and two orders of magnitude higher than Vakarchuk and Rovenchak \cite{Vakarchuk2001}. One can note from table \ref{tab:rholam} that $\rho \Lambda_c^3$ and $C_N^{\rm max}/k_B N$ are converging toward the expected values of $\rho \Lambda_c^3 = \zeta(3/2) = 2.612$ and $C_N^{\rm max}/k_B N = 1.926$ for the limit of $N \rightarrow \infty$ \cite{Vakarchuk2001}. 

\section{Discussion}  \label{sec-Discussion}
As mentioned in the Introduction, Weinberg highlights two unsatisfactory elements of non-relativistic quantum mechanics:  collapse of the wave function and entanglement. Both of these can be understood and visualized using the SCFT results for the ideal Bose gas. As just shown, ring polymer SCFT is able to reproduce the quantum phenomenon of a BEC exactly without using state vectors. While the quantum-classical isomorphism maps classical statistical mechanics onto quantum statistical mechanics, polymer SCFT goes further and maps classical statistical mechanics onto non-relativistic quantum mechanics through the theorems of DFT. Since SCFT is a statistical mechanics methodology, one is forced to picture quantum phenomena as ensembles, that is, from the SCFT perspective, quantum mechanics is not applicable to individual systems, only to ensembles of systems in the sense of Ballentine \cite{Ballentine1970, Bransden2000, Peres2002}, explaining the origins of quantum probabilities. The ``hidden variables'' of quantum mechanical microstates over which ensembles are calculated can then be visualized from the SCFT point-of-view as the conformations the ring polymer threads explore within the thermal-space, both in equilibrium and for time-dependent phenomena \cite{Thompson2022}. As discussed by Ballentine \cite{Ballentine1970}, there is no wave function collapse in an ensemble interpretation because it is understood that the wave function does not describe an individual system. This is made explicitly clear in the SCFT equations in that they do not contain wave functions and yet, through the theorems of DFT, they are guaranteed to make all the same predictions as wave function quantum mechanics \cite{Hohenberg1964, Mermin1965, Runge1984, Thompson2019, Thompson2022, Thompson2023}. 

In addition to avoiding the measurement problem by picturing quantum particles as ensembles of polymer conformations, these conformations allow one to visualize entanglement. A single quantum particle, when pictured as a ring polymer in the thermal-space, can be instantaneously affected by any change in a surrounding experimental setup. To understand this, note that a ring polymer is non-local, statistically extending an arbitrary distance in space as shown schematically in figure \ref{fig-exchange}. A change in the experimental setup, including a change to measurement settings, must be represented mathematically in the diffusion equation (\ref{diff1}) as a change in the external field contribution to $w({\bf r})$. One can ask how fast such a change touching one portion of the polymer trajectory might affect all other portions of the thread. An answer is that there is no ``speed limit'' in thermal-space; the imaginary time is space-like making the thermal-space Euclidean rather than Lorentzian, so a speed of light restriction does not apply at each moment in time within the thermal-space of a ring polymer. The speed of light limit remains, of course, along the space-time direction, but in the SCFT 5D description there is no mathematical justification for imposing an internal time delay for the effects of an external change of conditions on the results of a measurement for a single quantum particle. The effect can be instantaneous, although it is clearly not an action-at-a-distance since the polymer is correlated along its contour and this thermal connectivity propagates the effects of changing external fields. The SCFT perspective can therefore be considered a different way of looking at contextuality -- the results of a measurement depend on other measurements of compatible observables \cite{LaCour2009}. In SCFT, this occurs because the fields $w({\bf r})$ in equation (\ref{diff1}), which mathematically articulate the experimental setup, affect the polymer trajectories.\footnote{The SCFT perspective on contextuality does not conflict with the Kochen-Specker theorem because the polymer conformations are themselves contextual.}

The same argument holds for entangled sets of quantum particles if entanglement is identified with boson exchange. Simon has pointed out that the very nature of a BEC is that it is in a very highly entangled state \cite{Simon2002}. Although a BEC can be written in terms of product states based on the particle number $N$ of the condensate, Simon explains that particle numbers are not appropriate subsystems in the context of a BEC due to indistiguishability, but rather that spatial modes can be a natural physical choice for a discussion of entanglement in this system. The BEC entanglement proof of Simon is consistent with the views of other authors \cite{Laflorencie2016,Schmied2016,DelMaestro2020}, with Heaney and Vedral stating that the concept of \emph{particle} entanglement is meaningless in a BEC and that spatial modes are the natural description \cite{Heaney2009}. Since Bose-Einstein condensation and entanglement are both inherently quantum phenomena and, as derived in this paper and elsewhere, the only quantum feature needed to produce a BEC is boson exchange, then the quantum entanglement of the condensate must necessarily be mapped onto boson exchange: in a BEC, mode entanglement is boson exchange and vice versa.

This line of reasoning can be pushed further to particle (as opposed to mode) entanglement. Within the quantum-classical isomorphism, boson exchange is equivalent to two or more rings merging together to form a longer ring. The composite ring is identical in mathematical structure to a single quantum particle, now extended however over longer distances.\footnote{All rings in principle extend to infinity, but the probability of detection diminishes with distance according to the diffusion equation (\ref{diff1}). When multiple rings merge to form a longer ring, the probability of detection at a remote distance increases accordingly.} A composite ring in an experimental setup can potentially propagate any changes of apparatus settings through the thermal degree of freedom instantaneously, resulting in what is observed to be entanglement. This is the situation that, according to Bell, ``... is just what we would wish to avoid'' \cite{Bell1964}, in that it means a measurement at one device will now depend on the settings of a remote apparatus. The mathematics of the quantum-classical isomorphism coupled with the theorems of DFT through SCFT mean that this situation should be considered as a plausible description of nature as demonstrated by BEC systems with the proposed exchange-entanglement mapping. Note that this mechanism holds for multipartite and non-statistical entanglement such as GHZ states as well as bipartite entanglement.\footnote{It was previously speculated that the conformations of a polymer could be used to explain entanglement using distinct polymers of identical configurations \cite{Thompson2023}. However the non-locality of identical but distinct polymers is not sufficient to satisfy Bell-type non-locality, whereas the current picture of merged rings does satisfy it.} Similar ideas have been explored by Myrvold \cite{Myrvold2016} from a non-thermal perspective and the Wharton group from block-universe \cite{Wharton2016} and sum-over-histories \cite{Wharton2022, Wharton2023} path integral perspectives.

Since entanglement is inherent in a BEC, one might expect the entanglement entropy to obey an area law. Entropy often scales with the volume of a system, but a hallmark of entanglement is that the dominant term of the entanglement entropy scales instead with the area of the boundary of the volume \cite{Eisert2010, Laflorencie2016}. Non-interacting boson systems however, such as the ideal Bose gas, are exceptions. Laflorencie points out that for non-relativistic, non-interacting boson systems displaying Bose-Einstein condensation, the area law term is suppressed and the dominant term is logarithmic in the number of particles \cite{Laflorencie2016}. SCFT can be seen to agree with this finding. For an ideal Bose gas, both the potential $U$ and field $w({\bf r})$ are zero in the free energy expression (\ref{FE1}) giving
\begin{equation}
F = -\frac{N}{\beta} \ln Q_1  \label{FE2}
\end{equation}
where the factorial term in (\ref{FE1}) has been dropped for convenience. Dropping the factorial term means that the zero of free energy is taken as the classical uniform gas, and so all contributions to (\ref{FE2}) are from the configurational entropy which accounts for the conformations available to the ring polymer. This is a good entropic measure for quantum behaviour since the classical conformations of ring polymers map onto the quantum nature of the system in the quantum-classical isomorphism. To include boson exchange, $Q_1$ should be replaced with $Q_N$ in equation (\ref{FE2}) giving a configurational entropy of 
\begin{equation}
S_c = N \ln Q_N .   \label{Sc1}
\end{equation}
Equation (\ref{QN}) can be rewritten as
\begin{equation}
Q_{N}(\beta) = \sum_{j=1}^{P(N)} c_j \prod_{k \in \delta_j} k^{\frac{3}{2}g(k)}Q_1({k\beta})^{g(k)}   \label{QN2}
\end{equation}
and so $Q_N$ for a perfect BEC, where all $N$ rings have combined into a single ring, can be seen to be just
\begin{equation}
Q_{N}(\beta) = Q_1(N\beta) = N^{-\frac{3}{2}} Q_1(\beta)   \label{QN3}
\end{equation}
where equation (\ref{Q1b}) has been used for the second equality. For this case, the prefactor $N$ in equation (\ref{Sc1}) should be set to one, since there is only one large ring polymer of length $N\beta$, and the configurational entropy for a perfect BEC is therefore
\begin{equation}
S_c = -\frac{3}{2} \ln N + \ln\left(\frac{V}{\Lambda^3}\right)    \label{Sc2}
\end{equation}
which scales to leading order logarithmically with number of fundamental rings $N$ that make up the single BEC ring. This agrees with previous results for entanglement entropy in ideal Bose gases using other methods \cite{Alba2013, Ding2009, Simon2002}.

\section{Summary and Future Outlook} 
The specific heats and critical temperatures for the ideal Bose gas have been calculated using SCFT for an order of magnitude more particles than has previously been reported. The governing diffusion equation of SCFT was trivially solved for the uniform ideal gas with boson exchange giving the Matsubara expressions (\ref{QN}) and (\ref{cj}) \cite{Mullin2000, Matsubara1951, Feynman1972}, guaranteeing equivalence of predictions between ring polymer SCFT and standard quantum mechanics. This is as required, since ring polymer SCFT has previously been shown to be equivalent to quantum DFT \cite{Thompson2019} which, through the theorems of DFT \cite{Hohenberg1964, Mermin1965, Runge1984}, is itself equivalent to wave function non-relativistic quantum mechanics.

The SCFT results shown in figure \ref{fig-heat} confirm that the Bose gas undergoes a $\lambda$-transition and behaves as a BEC. Given that the only quantum feature of the ideal Bose BEC system is boson exchange, and applying the quantum field theory proof of Simon that the nature of a BEC arises solely from mode entanglement, it is proven by transitivity that mode entanglement in the ideal BEC is equivalent to boson exchange. One can consider whether this exchange-entanglement mapping is universal beyond this system, and the ring polymer perspective gives insight into this. Since boson exchange can universally be viewed in ring polymer quantum theory as the merging and separation of rings, it is possible to visualize entanglement in more general systems and confirm behaviours are as expected. As discussed in Section \ref{sec-Discussion}, this ring polymer picture allows an easy visualization of bipartite and multipartite particle entanglement for arbitrary systems, including non-statistical entanglement, and gives an intuitive way of viewing contextuality.

Since the SCFT equations do not involve state vectors, there is no collapse of the wave function in ring polymer quantum theory. This can be understood more generally in that SCFT is a statistical mechanics theory based on ensembles, and so the measurement problem is avoided following the ensemble interpretation expounded by Ballentine \cite{Ballentine1970}. Instead of state vectors, a density matrix which is the solution of the diffusion equation (\ref{diff1}) is interpreted physically as quantifiying ring polymer conformations in thermal-space. Ring polymer SCFT therefore allows one to visualize both entanglement and the behaviour of individual quantum systems.

Important future work could include exploring systems beyond the ideal Bose gas. While a non-interacting Bose gas does not show an entanglement entropy area law, Laflorencie comments that any interactions in a BEC, such as those giving rise to superfluidity, should give rise to an area law \cite{Laflorencie2016}. Fredrickson et al. have already considered interacting Bose gases using FTS \cite{Fredrickson2020, Fredrickson2022}, so it is a logical step to attempt incorporating interactions in SCFT. FTS is closely related to SCFT but goes beyond a mean field description by including fluctuations. On the other hand, SCFT is much simpler theoretically and computationally, so it may be worth exploring whether an SCFT description of an interacting Bose system might give the expected area law, or if the mean field nature precludes this. Qualitatively, the SCFT/FTS descriptions should give entanglement entropy area laws, since the ring polymer system can be mapped onto a classical system that is itself known to obey an area law. Specifically, Eisert has proven that a classical lattice system of coupled harmonic oscillators obeys an area law for the mutual information, analogous to the entanglement entropy \cite{Eisert2010}. The merged ring polymers of the quantum-classical isomorphism appear in thermal-space as masses coupled by harmonic springs, the bare polymer Gaussian thread model of SCFT having the same mathematical form as a spring potential -- see the effective Hamiltonian term $\frac{m}{2\hbar^2} \left(\frac{d{\bf r}}{ds}\right)^2$ in equation (\ref{part1}). This is isomorphic to the classical harmonic system described by Eisert and so should obey an area law. An explicit SCFT calculation of a superfluid system would nonetheless be valuable to either verify this result or to limit the applicability of the mean field approximation of SCFT. Due to the correspondence between SCFT and DFT, some correlation corrections to the mean field could be included in an exchange-correlation functional just as in DFT. While SCFT and DFT are in principle equivalent, SCFT does have some scope for exchange-correlation corrections that may not be obvious in DFT, such as extensions to FTS and including exact exchange effects as done in this paper for boson exchange and in reference \onlinecite{Kealey2024} for fermion exchange. 

One might also ask how this model of boson entanglement, visualized as the merging of ring polymers, fits with the entanglement of fermions. For bipartite entanglement, there is no problem since fermions of differing spin can occupy the same state and be treated in SCFT as merging rings also. This has been verified numerically for the case of the helium atom \cite{Thompson2023}. It is not clear how to proceed for multipartite fermion entanglement. One possibility is to model this case as the merging of composite bosons. For example, two opposite spin electrons could merge (entangle) to form a composite boson, and multiple such composite bosons could be pictured as merging with each other or with single electrons to form larger multipartite rings. This speculative model for fermion entanglement bears some superficial resemblance to multi-step entanglement processes used in some quantum computing architectures involving electron GHZ states \cite{Baugh2019}. It also follows the spirit of Feynman's original application of ring polymer quantum theory to helium atoms, which are composite particles built out of fermions \cite{Feynman1953b}. Hierarchical fermion entanglement of this sort could be considered in future work.

\appendix
\section{Ring Polymer Combinatorial Coefficients from Integer Partitions }
The expression (\ref{QN})-(\ref{cj}) can be derived by combinatorial analysis of the different ways of merging $N$ ring polymers using the integer partition \cite{Mullin2000}. A ring polymer loop of contour length $\beta$ has partition function $Q_1(\beta)$. So, for example, four loops, $N=4$, can be merged into one large loop of length $4\beta$ with partition function $Q_1(4\beta)$, or a loop of $3\beta$ and a loop of $\beta$ with partition functions $Q_1(3\beta)$ and $Q_1(\beta)$, respectively, or $Q_1(2\beta)$ repeated twice, or $Q_1(\beta)$ repeated four times. The expression for $Q_N(\beta)$ then must be
\begin{equation}
Q_N(\beta) = \sum_{j=1}^{P(N)} \tilde{c}_j \prod_{k \in \varepsilon_j} Q(k\beta)  \label{QN4}
\end{equation}
where $\tilde{c}_j$ are as of yet unspecified coefficients and $P(N)$ is the number of integer partitions of $N$. For example, if $N=4$, $P(4) = 5$ as explained in section \ref{sec-Theory}. The product is over the elements of each integer partition $\varepsilon_j$, so for example, with $N=4$, if $\varepsilon_1 = [1,1,1,1]$ then the product in (\ref{QN4}) gives $Q(\beta)^4$, whereas $\varepsilon_2 = [1,1,2]$ gives a term in the sum of $Q(\beta)^2Q(2\beta)$. $\varepsilon_j$ is a multi-set since it can have repeated elements, but using a degeneracy $g(k)$ defined in section \ref{sec-Theory}, equation (\ref{QN4}) can be rewritten as 
\begin{equation}
Q_N(\beta) = \sum_{j=1}^{P(N)} \tilde{c}_j Q(\beta)^{l_j} \prod_{k \in \delta_j} k^{-\frac{3}{2} g(k)}  \label{QN5}
\end{equation}
where the product $k$ is over the elements of the $j$th set $\delta_j$ of the integer partition with degeneracies $g(k)$, and $l_j$ is the cardinality of  $\varepsilon_j$ (or $\delta_j$ including degeneracy). Equation (\ref{QN5}) has been simplified using expressions (\ref{diff3}) and (\ref{Q1b}):
\begin{equation}
Q(k\beta) =  \frac{V}{\Lambda (k\beta)^3}  = k^{-\frac{3}{2}} \frac{V}{\Lambda (\beta)} = k^{-\frac{3}{2}} Q(\beta)    .  \label{Q1c}
\end{equation}

The coefficients $\tilde{c}_j$ of equations (\ref{QN4}) or (\ref{QN5}) can be found by counting the various $Q(k\beta)$ loops. The prefactors $\tilde{c}_j$ are then the products of the prefactors of each $Q(k\beta)$ for each term of the sum over $j$. The number of ways that one can build a ring of contour length $k_1\beta$ with partition function $Q(k_1 \beta)$ out of the $N$ polymers is
\begin{equation}
{N \choose k_1} = \frac{N!}{k_1!(N-k_1)!}  . \label{comb1}
\end{equation}
Subsequent loops of length $k_2\beta$ with partition function $Q(k_2 \beta)$ would be built out of the remaining $N-k_1$ rings, so that the product for each term of the sum over $j$ would be
\begin{equation}
\frac{N!}{k_1!(N-k_1)!} \times \frac{(N-k_1)!}{k_1!(N-k_1-k_2)!} \cdots = \frac{N!}{\prod_i k_i!(N-k_1-k_2 - \cdots)!}   \label{comb2}
\end{equation}
where (\ref{comb2}) is generalized for many rings sizes, $k_1, k_2, k_3, \cdots$. Since $N = k_1+k_2 + \cdots$, (\ref{comb2}) simplifies to just $N!/\prod_i k_i!$. There are multiple ways however to merge loops into larger rings. To build a ring of length $3\beta$ for example, one could add the fundamental rings of length $\beta$ in various orders: ring1+ring2+ring3 is a different configuration than ring1+ring3+ring2. Accounting for this contributes a factor $(k_i-1)!$ where the minus 1 appears because the starting point of the larger loop doesn't matter; all loops are cyclic. The prefactor for each $j$ is then 
\begin{equation}
\tilde{c}_j = \frac{N! (k-1)!}{\prod_{k \in \varepsilon_j} k!}  = \frac{N!}{\prod_{k \in \varepsilon_j} k}    \label{comb3}
\end{equation}
where equation (\ref{comb3}) is now written over the elements of the integer partition, rather than the subscript $i$,  to select the possible lengths $k$ of rings. This can be written instead over the sets $\delta_j$ with degeneracies rather than the multi-sets $\varepsilon_j$:
\begin{equation}
\tilde{c}_j = \frac{N!}{\prod_{k \in \delta_j} k^{g(k)}}  .  \label{comb4}
\end{equation}
Finally, it is necessary to divide expression (\ref{comb4}) by the factorial of the degeneracy to correct for double counting. The coefficients of equations (\ref{QN4}) or (\ref{QN5}) are therefore
\begin{equation}
\tilde{c}_j = \frac{N!}{\prod_{k \in \delta_j} k^{g(k)} g(k)!}  .  \label{comb5}
\end{equation}
From (\ref{QN5}), a factor of $k^{-\frac{3}{2}}$ can be pulled into these coefficients to instead write the Matsubara formula as 
\begin{equation}
Q_{N}(\beta) = \sum_{j=1}^{P(N)} c_j Q_{1}(\beta)^{l_j}  \label{QN6}
\end{equation}
with 
\begin{equation}
c_j = \frac{N!}{\prod_{k \in \delta_j} k^{\frac{5}{2} g(k)} g(k)!}  . \label{comb6}
\end{equation} 
as given by equations (\ref{QN}) and (\ref{cj}) in Section \ref{sec-Theory}.

\begin{acknowledgments}
The author acknowledges productive discussions with many participants of the sixteenth biennial meeting of the International Quantum Structures Association (IQSA) at Vrije Universiteit Brussel, and helpful communications with both Ken Wharton about his path integral techniques and Jonathan Baugh regarding GHZ states in quantum computing architectures.
\end{acknowledgments}

\bibliography{DFTbibliography}

\begin{thebibliography}{68}%
\makeatletter
\providecommand \@ifxundefined [1]{%
 \@ifx{#1\undefined}
}%
\providecommand \@ifnum [1]{%
 \ifnum #1\expandafter \@firstoftwo
 \else \expandafter \@secondoftwo
 \fi
}%
\providecommand \@ifx [1]{%
 \ifx #1\expandafter \@firstoftwo
 \else \expandafter \@secondoftwo
 \fi
}%
\providecommand \natexlab [1]{#1}%
\providecommand \enquote  [1]{``#1''}%
\providecommand \bibnamefont  [1]{#1}%
\providecommand \bibfnamefont [1]{#1}%
\providecommand \citenamefont [1]{#1}%
\providecommand \href@noop [0]{\@secondoftwo}%
\providecommand \href [0]{\begingroup \@sanitize@url \@href}%
\providecommand \@href[1]{\@@startlink{#1}\@@href}%
\providecommand \@@href[1]{\endgroup#1\@@endlink}%
\providecommand \@sanitize@url [0]{\catcode `\\12\catcode `\$12\catcode
  `\&12\catcode `\#12\catcode `\^12\catcode `\_12\catcode `\%12\relax}%
\providecommand \@@startlink[1]{}%
\providecommand \@@endlink[0]{}%
\providecommand \url  [0]{\begingroup\@sanitize@url \@url }%
\providecommand \@url [1]{\endgroup\@href {#1}{\urlprefix }}%
\providecommand \urlprefix  [0]{URL }%
\providecommand \Eprint [0]{\href }%
\providecommand \doibase [0]{http://dx.doi.org/}%
\providecommand \selectlanguage [0]{\@gobble}%
\providecommand \bibinfo  [0]{\@secondoftwo}%
\providecommand \bibfield  [0]{\@secondoftwo}%
\providecommand \translation [1]{[#1]}%
\providecommand \BibitemOpen [0]{}%
\providecommand \bibitemStop [0]{}%
\providecommand \bibitemNoStop [0]{.\EOS\space}%
\providecommand \EOS [0]{\spacefactor3000\relax}%
\providecommand \BibitemShut  [1]{\csname bibitem#1\endcsname}%
\let\auto@bib@innerbib\@empty
\bibitem [{\citenamefont {Bransden}\ and\ \citenamefont
  {Joachain}(2003)}]{Bransden2000}%
  \BibitemOpen
  \bibfield  {author} {\bibinfo {author} {\bibfnamefont {B.~H.}\ \bibnamefont
  {Bransden}}\ and\ \bibinfo {author} {\bibfnamefont {C.~J.}\ \bibnamefont
  {Joachain}},\ }\href@noop {} {\emph {\bibinfo {title} {Quantum Mechanics}}}\
  (\bibinfo  {publisher} {Pearson},\ \bibinfo {address} {Dorchester, UK},\
  \bibinfo {year} {2003})\BibitemShut {NoStop}%
\bibitem [{\citenamefont {Cohen-{T}annoudji}\ \emph {et~al.}(1977)\citenamefont
  {Cohen-{T}annoudji}, \citenamefont {Diu},\ and\ \citenamefont
  {Lalo\"{e}}}]{CohenTannoudji1977}%
  \BibitemOpen
  \bibfield  {author} {\bibinfo {author} {\bibfnamefont {C.}~\bibnamefont
  {Cohen-{T}annoudji}}, \bibinfo {author} {\bibfnamefont {B.}~\bibnamefont
  {Diu}}, \ and\ \bibinfo {author} {\bibfnamefont {F.}~\bibnamefont
  {Lalo\"{e}}},\ }\href@noop {} {\emph {\bibinfo {title} {Quantum Mechanics}}}\
  (\bibinfo  {publisher} {John Wiley \& Sons},\ \bibinfo {address} {Paris,
  France},\ \bibinfo {year} {1977})\BibitemShut {NoStop}%
\bibitem [{\citenamefont {Shankar}(1994)}]{Shankar1994}%
  \BibitemOpen
  \bibfield  {author} {\bibinfo {author} {\bibfnamefont {R.}~\bibnamefont
  {Shankar}},\ }\href@noop {} {\emph {\bibinfo {title} {Principles of Quantum
  Mechanics}}}\ (\bibinfo  {publisher} {Springer},\ \bibinfo {address} {New
  York, USA},\ \bibinfo {year} {1994})\BibitemShut {NoStop}%
\bibitem [{\citenamefont {Bassi}\ and\ \citenamefont
  {Ghirardi}(2003)}]{Bassi2003}%
  \BibitemOpen
  \bibfield  {author} {\bibinfo {author} {\bibfnamefont {Angelo}\ \bibnamefont
  {Bassi}}\ and\ \bibinfo {author} {\bibfnamefont {GianCarlo}\ \bibnamefont
  {Ghirardi}},\ }\bibfield  {title} {\enquote {\bibinfo {title} {Dynamical
  reduction models},}\ }\href@noop {} {\bibfield  {journal} {\bibinfo
  {journal} {Physics Reports}\ }\textbf {\bibinfo {volume} {379}},\ \bibinfo
  {pages} {257--426} (\bibinfo {year} {2003})}\BibitemShut {NoStop}%
\bibitem [{\citenamefont {Vaidman}(2022)}]{Vaidman2022}%
  \BibitemOpen
  \bibfield  {author} {\bibinfo {author} {\bibfnamefont {Lev}\ \bibnamefont
  {Vaidman}},\ }\bibfield  {title} {\enquote {\bibinfo {title} {Why the
  many-worlds interpretation?}}\ }\href@noop {} {\bibfield  {journal} {\bibinfo
   {journal} {Quantum Reports}\ }\textbf {\bibinfo {volume} {4}},\ \bibinfo
  {pages} {264--271} (\bibinfo {year} {2022})}\BibitemShut {NoStop}%
\bibitem [{\citenamefont {Goldstein}\ \emph {et~al.}(2011)\citenamefont
  {Goldstein}, \citenamefont {Tumulka},\ and\ \citenamefont
  {Zanghì}}]{Goldstein2010}%
  \BibitemOpen
  \bibfield  {author} {\bibinfo {author} {\bibfnamefont {Sheldon}\ \bibnamefont
  {Goldstein}}, \bibinfo {author} {\bibfnamefont {Roderich}\ \bibnamefont
  {Tumulka}}, \ and\ \bibinfo {author} {\bibfnamefont {Nino}\ \bibnamefont
  {Zanghì}},\ }\bibfield  {title} {\enquote {\bibinfo {title} {Bohmian
  trajectories as the foundation of quantum mechanics},}\ }in\ \href@noop {}
  {\emph {\bibinfo {booktitle} {{Q}uantum {T}rajectories}}},\ \bibinfo {editor}
  {edited by\ \bibinfo {editor} {\bibfnamefont {Pratim~Kumar}\ \bibnamefont
  {Chattaraj}}}\ (\bibinfo  {publisher} {CRC Press, Taylor \& Francis},\
  \bibinfo {address} {New York},\ \bibinfo {year} {2011})\ pp.\ \bibinfo
  {pages} {1--16}\BibitemShut {NoStop}%
\bibitem [{\citenamefont {Weinberg}(2014)}]{Weinberg2014}%
  \BibitemOpen
  \bibfield  {author} {\bibinfo {author} {\bibfnamefont {Steven}\ \bibnamefont
  {Weinberg}},\ }\bibfield  {title} {\enquote {\bibinfo {title} {Quantum
  mechanics without state vectors},}\ }\href@noop {} {\bibfield  {journal}
  {\bibinfo  {journal} {Physical Review A}\ }\textbf {\bibinfo {volume} {90}},\
  \bibinfo {pages} {042102} (\bibinfo {year} {2014})}\BibitemShut {NoStop}%
\bibitem [{Wei()}]{Weinberg2014pre}%
  \BibitemOpen
  \href@noop {} {}\Eprint {http://arxiv.org/abs/1405.3483} {arXiv:1405.3483
  [quant-ph]} \BibitemShut {NoStop}%
\bibitem [{\citenamefont {Feynman}\ and\ \citenamefont
  {Hibbs}(1965)}]{Feynman1965}%
  \BibitemOpen
  \bibfield  {author} {\bibinfo {author} {\bibfnamefont {Richard~P.}\
  \bibnamefont {Feynman}}\ and\ \bibinfo {author} {\bibfnamefont {Albert~R.}\
  \bibnamefont {Hibbs}},\ }\href@noop {} {\emph {\bibinfo {title} {Quantum
  Mechanics and Path Integrals}}}\ (\bibinfo  {publisher} {Dover
  Publications},\ \bibinfo {address} {Mineola NY},\ \bibinfo {year}
  {1965})\BibitemShut {NoStop}%
\bibitem [{\citenamefont {Ceperley}(1995)}]{Ceperley1995}%
  \BibitemOpen
  \bibfield  {author} {\bibinfo {author} {\bibfnamefont {D.~M.}\ \bibnamefont
  {Ceperley}},\ }\bibfield  {title} {\enquote {\bibinfo {title} {Path integrals
  in the theory of condensed helium},}\ }\href@noop {} {\bibfield  {journal}
  {\bibinfo  {journal} {Reviews of Modern Physics}\ }\textbf {\bibinfo {volume}
  {67}},\ \bibinfo {pages} {279--355} (\bibinfo {year} {1995})}\BibitemShut
  {NoStop}%
\bibitem [{\citenamefont {Feynman}(1953)}]{Feynman1953b}%
  \BibitemOpen
  \bibfield  {author} {\bibinfo {author} {\bibfnamefont {Richard~P.}\
  \bibnamefont {Feynman}},\ }\bibfield  {title} {\enquote {\bibinfo {title}
  {Atomic theory of the $\lambda$-transition in helium},}\ }\href@noop {}
  {\bibfield  {journal} {\bibinfo  {journal} {Physical Review}\ }\textbf
  {\bibinfo {volume} {91}},\ \bibinfo {pages} {1291--1301} (\bibinfo {year}
  {1953})}\BibitemShut {NoStop}%
\bibitem [{\citenamefont {Roy}\ and\ \citenamefont {Voth}(1999)}]{Roy1999b}%
  \BibitemOpen
  \bibfield  {author} {\bibinfo {author} {\bibfnamefont {Pierre-Nicholas}\
  \bibnamefont {Roy}}\ and\ \bibinfo {author} {\bibfnamefont {G.~A.}\
  \bibnamefont {Voth}},\ }\bibfield  {title} {\enquote {\bibinfo {title}
  {Feynman path centroid dynamics for {Fermi-Dirac} statistics},}\ }\href@noop
  {} {\bibfield  {journal} {\bibinfo  {journal} {J. Chem. Phys.}\ }\textbf
  {\bibinfo {volume} {111}},\ \bibinfo {pages} {5303--5305} (\bibinfo {year}
  {1999})}\BibitemShut {NoStop}%
\bibitem [{\citenamefont {Zeng}\ and\ \citenamefont {Roy}(2014)}]{Zeng2014}%
  \BibitemOpen
  \bibfield  {author} {\bibinfo {author} {\bibfnamefont {Tao}\ \bibnamefont
  {Zeng}}\ and\ \bibinfo {author} {\bibfnamefont {Pierre-Nicholas}\
  \bibnamefont {Roy}},\ }\bibfield  {title} {\enquote {\bibinfo {title}
  {Microscopic molecular superfluid response: theory and simulations},}\
  }\href@noop {} {\bibfield  {journal} {\bibinfo  {journal} {Reports on
  Progress in Physics}\ }\textbf {\bibinfo {volume} {77}},\ \bibinfo {pages}
  {046601} (\bibinfo {year} {2014})}\BibitemShut {NoStop}%
\bibitem [{\citenamefont {Richardson}\ and\ \citenamefont
  {Althorpe}(2009)}]{Althorpe2009}%
  \BibitemOpen
  \bibfield  {author} {\bibinfo {author} {\bibfnamefont {Jeremy~O.}\
  \bibnamefont {Richardson}}\ and\ \bibinfo {author} {\bibfnamefont
  {Stuart~C.}\ \bibnamefont {Althorpe}},\ }\bibfield  {title} {\enquote
  {\bibinfo {title} {Ring-polymer molecular dynamics rate-theory in the
  deep-tunneling regime: Connection with semiclassical instanton theory},}\
  }\href@noop {} {\bibfield  {journal} {\bibinfo  {journal} {Journal of
  Chemical Physics}\ }\textbf {\bibinfo {volume} {131}},\ \bibinfo {pages}
  {214106} (\bibinfo {year} {2009})}\BibitemShut {NoStop}%
\bibitem [{\citenamefont {Habershon}\ \emph {et~al.}(2013)\citenamefont
  {Habershon}, \citenamefont {Manolopoulos}, \citenamefont {Markland},\ and\
  \citenamefont {Miller}}]{Habershon2013}%
  \BibitemOpen
  \bibfield  {author} {\bibinfo {author} {\bibfnamefont {Scott}\ \bibnamefont
  {Habershon}}, \bibinfo {author} {\bibfnamefont {David~E.}\ \bibnamefont
  {Manolopoulos}}, \bibinfo {author} {\bibfnamefont {Thomas~E.}\ \bibnamefont
  {Markland}}, \ and\ \bibinfo {author} {\bibfnamefont {Thomas~F.}\
  \bibnamefont {Miller}},\ }\bibfield  {title} {\enquote {\bibinfo {title}
  {Ring-polymer molecular dynamics: Quantum effects in chemical dynamics from
  classical trajectories in an extended phase space},}\ }\href@noop {}
  {\bibfield  {journal} {\bibinfo  {journal} {Annual Review of Physical
  Chemistry}\ }\textbf {\bibinfo {volume} {64}},\ \bibinfo {pages} {387--413}
  (\bibinfo {year} {2013})}\BibitemShut {NoStop}%
\bibitem [{\citenamefont {Thompson}(2019)}]{Thompson2019}%
  \BibitemOpen
  \bibfield  {author} {\bibinfo {author} {\bibfnamefont {Russell~B.}\
  \bibnamefont {Thompson}},\ }\bibfield  {title} {\enquote {\bibinfo {title}
  {An alternative derivation of orbital-free density functional theory},}\
  }\href@noop {} {\bibfield  {journal} {\bibinfo  {journal} {Journal of
  Chemical Physics}\ }\textbf {\bibinfo {volume} {150}},\ \bibinfo {pages}
  {204109} (\bibinfo {year} {2019})}\BibitemShut {NoStop}%
\bibitem [{\citenamefont {Thompson}(2020)}]{Thompson2020}%
  \BibitemOpen
  \bibfield  {author} {\bibinfo {author} {\bibfnamefont {Russell~B.}\
  \bibnamefont {Thompson}},\ }\bibfield  {title} {\enquote {\bibinfo {title}
  {Atomic shell structure from an orbital-free-related
  density-functional-theory {Pauli} potential},}\ }\href@noop {} {\bibfield
  {journal} {\bibinfo  {journal} {Physical Review A}\ }\textbf {\bibinfo
  {volume} {102}},\ \bibinfo {pages} {012813} (\bibinfo {year}
  {2020})}\BibitemShut {NoStop}%
\bibitem [{\citenamefont {Sillaste}\ and\ \citenamefont
  {Thompson}(2022)}]{Sillaste2022}%
  \BibitemOpen
  \bibfield  {author} {\bibinfo {author} {\bibfnamefont {Spencer}\ \bibnamefont
  {Sillaste}}\ and\ \bibinfo {author} {\bibfnamefont {Russell~B.}\ \bibnamefont
  {Thompson}},\ }\bibfield  {title} {\enquote {\bibinfo {title} {Molecular
  bonding in an orbital-free-related density functional theory},}\ }\href@noop
  {} {\bibfield  {journal} {\bibinfo  {journal} {Journal of Physical Chemistry
  A}\ }\textbf {\bibinfo {volume} {126}},\ \bibinfo {pages} {325--332}
  (\bibinfo {year} {2022})}\BibitemShut {NoStop}%
\bibitem [{\citenamefont {Thompson}(2022)}]{Thompson2022}%
  \BibitemOpen
  \bibfield  {author} {\bibinfo {author} {\bibfnamefont {Russell~B.}\
  \bibnamefont {Thompson}},\ }\bibfield  {title} {\enquote {\bibinfo {title}
  {An interpretation of quantum foundations based on density functional theory
  and polymer self-consistent field theory},}\ }\href@noop {} {\bibfield
  {journal} {\bibinfo  {journal} {Quantum Studies: Mathematics and
  Foundations}\ }\textbf {\bibinfo {volume} {9}},\ \bibinfo {pages} {405--416}
  (\bibinfo {year} {2022})}\BibitemShut {NoStop}%
\bibitem [{\citenamefont {Thompson}(2023)}]{Thompson2023}%
  \BibitemOpen
  \bibfield  {author} {\bibinfo {author} {\bibfnamefont {Russell~B.}\
  \bibnamefont {Thompson}},\ }\bibfield  {title} {\enquote {\bibinfo {title} {A
  holographic principle for non-relativistic quantum mechanics},}\ }\href@noop
  {} {\bibfield  {journal} {\bibinfo  {journal} {International Journal of
  Theoretical Physics}\ }\textbf {\bibinfo {volume} {62}},\ \bibinfo {pages}
  {34:1--15} (\bibinfo {year} {2023})}\BibitemShut {NoStop}%
\bibitem [{\citenamefont {LeMaitre}\ and\ \citenamefont
  {Thompson}(2023{\natexlab{a}})}]{LeMaitre2023a}%
  \BibitemOpen
  \bibfield  {author} {\bibinfo {author} {\bibfnamefont {Phil~A.}\ \bibnamefont
  {LeMaitre}}\ and\ \bibinfo {author} {\bibfnamefont {Russell~B.}\ \bibnamefont
  {Thompson}},\ }\bibfield  {title} {\enquote {\bibinfo {title} {Gaussian basis
  functions for an orbital-free-related density functional theory of atoms},}\
  }\href@noop {} {\bibfield  {journal} {\bibinfo  {journal} {International
  Journal of Quantum Chemistry}\ }\textbf {\bibinfo {volume} {123}},\ \bibinfo
  {pages} {e27111} (\bibinfo {year} {2023}{\natexlab{a}})}\BibitemShut
  {NoStop}%
\bibitem [{\citenamefont {LeMaitre}\ and\ \citenamefont
  {Thompson}(2023{\natexlab{b}})}]{LeMaitre2023b}%
  \BibitemOpen
  \bibfield  {author} {\bibinfo {author} {\bibfnamefont {Phil~A.}\ \bibnamefont
  {LeMaitre}}\ and\ \bibinfo {author} {\bibfnamefont {Russell~B.}\ \bibnamefont
  {Thompson}},\ }\bibfield  {title} {\enquote {\bibinfo {title} {On the origins
  of spontaneous spherical symmetry-breaking in open-shell atoms through
  polymer self-consistent field theory},}\ }\href@noop {} {\bibfield  {journal}
  {\bibinfo  {journal} {Journal of Chemical Physics}\ }\textbf {\bibinfo
  {volume} {158}},\ \bibinfo {pages} {064301} (\bibinfo {year}
  {2023}{\natexlab{b}})}\BibitemShut {NoStop}%
\bibitem [{\citenamefont {Kealey}\ \emph {et~al.}(2024)\citenamefont {Kealey},
  \citenamefont {LeMaitre},\ and\ \citenamefont {Thompson}}]{Kealey2024}%
  \BibitemOpen
  \bibfield  {author} {\bibinfo {author} {\bibfnamefont {Malcolm~A.}\
  \bibnamefont {Kealey}}, \bibinfo {author} {\bibfnamefont {Phil~A.}\
  \bibnamefont {LeMaitre}}, \ and\ \bibinfo {author} {\bibfnamefont
  {Russell~B.}\ \bibnamefont {Thompson}},\ }\bibfield  {title} {\enquote
  {\bibinfo {title} {Fermion exchange in ring polymer quantum theory},}\
  }\href@noop {} {\bibfield  {journal} {\bibinfo  {journal} {Physical Review
  A}\ }\textbf {\bibinfo {volume} {109}},\ \bibinfo {pages} {052819} (\bibinfo
  {year} {2024})}\BibitemShut {NoStop}%
\bibitem [{\citenamefont {Hohenberg}\ and\ \citenamefont
  {Kohn}(1964)}]{Hohenberg1964}%
  \BibitemOpen
  \bibfield  {author} {\bibinfo {author} {\bibfnamefont {P.}~\bibnamefont
  {Hohenberg}}\ and\ \bibinfo {author} {\bibfnamefont {W.}~\bibnamefont
  {Kohn}},\ }\bibfield  {title} {\enquote {\bibinfo {title} {Inhomogeneous
  electron gas},}\ }\href@noop {} {\bibfield  {journal} {\bibinfo  {journal}
  {Physical Review}\ }\textbf {\bibinfo {volume} {136}},\ \bibinfo {pages}
  {B864--B871} (\bibinfo {year} {1964})}\BibitemShut {NoStop}%
\bibitem [{\citenamefont {Chandler}\ and\ \citenamefont
  {Wolynes}(1981)}]{Chandler1981}%
  \BibitemOpen
  \bibfield  {author} {\bibinfo {author} {\bibfnamefont {David}\ \bibnamefont
  {Chandler}}\ and\ \bibinfo {author} {\bibfnamefont {P.~W.}\ \bibnamefont
  {Wolynes}},\ }\bibfield  {title} {\enquote {\bibinfo {title} {Exploiting the
  isomorphism between quantum theory and classical statistical mechanics of
  polyatomic fluids},}\ }\href@noop {} {\bibfield  {journal} {\bibinfo
  {journal} {Journal of Chemical Physics}\ }\textbf {\bibinfo {volume} {74}},\
  \bibinfo {pages} {4078--4095} (\bibinfo {year} {1981})}\BibitemShut {NoStop}%
\bibitem [{\citenamefont {Mermin}(1965)}]{Mermin1965}%
  \BibitemOpen
  \bibfield  {author} {\bibinfo {author} {\bibfnamefont {N.~David}\
  \bibnamefont {Mermin}},\ }\bibfield  {title} {\enquote {\bibinfo {title}
  {Thermal properties of the inhomogeneous electron gas},}\ }\href@noop {}
  {\bibfield  {journal} {\bibinfo  {journal} {Physical Review}\ }\textbf
  {\bibinfo {volume} {137}},\ \bibinfo {pages} {A1441--A1443} (\bibinfo {year}
  {1965})}\BibitemShut {NoStop}%
\bibitem [{\citenamefont {Runge}\ and\ \citenamefont
  {Gross}(1984)}]{Runge1984}%
  \BibitemOpen
  \bibfield  {author} {\bibinfo {author} {\bibfnamefont {Erich}\ \bibnamefont
  {Runge}}\ and\ \bibinfo {author} {\bibfnamefont {Eberhard K.~U.}\
  \bibnamefont {Gross}},\ }\bibfield  {title} {\enquote {\bibinfo {title}
  {Density-functional theory for time-dependent systems},}\ }\href@noop {}
  {\bibfield  {journal} {\bibinfo  {journal} {Phys. Rev. Lett.}\ }\textbf
  {\bibinfo {volume} {52}},\ \bibinfo {pages} {997--1000} (\bibinfo {year}
  {1984})}\BibitemShut {NoStop}%
\bibitem [{\citenamefont {Kaluza}(1921)}]{Kaluza1921}%
  \BibitemOpen
  \bibfield  {author} {\bibinfo {author} {\bibfnamefont {Theodor}\ \bibnamefont
  {Kaluza}},\ }\bibfield  {title} {\enquote {\bibinfo {title} {Zum
  unit\"{a}tsproblem der physik},}\ }\href@noop {} {\bibfield  {journal}
  {\bibinfo  {journal} {Sitz. Preuss. Akad. Wiss. Phys. Math. Kl.}\ ,\ \bibinfo
  {pages} {966--972}} (\bibinfo {year} {1921})}\BibitemShut {NoStop}%
\bibitem [{\citenamefont {Overduin}\ and\ \citenamefont
  {Wesson}(1997)}]{Wesson1997}%
  \BibitemOpen
  \bibfield  {author} {\bibinfo {author} {\bibfnamefont {James~M.}\
  \bibnamefont {Overduin}}\ and\ \bibinfo {author} {\bibfnamefont {Paul~S.}\
  \bibnamefont {Wesson}},\ }\bibfield  {title} {\enquote {\bibinfo {title}
  {{K}aluza-{K}lein gravity},}\ }\href@noop {} {\bibfield  {journal} {\bibinfo
  {journal} {Phys. Rep.}\ }\textbf {\bibinfo {volume} {283}},\ \bibinfo {pages}
  {303--378} (\bibinfo {year} {1997})}\BibitemShut {NoStop}%
\bibitem [{\citenamefont {Dinov}\ and\ \citenamefont
  {Velev}(2022)}]{Dinov2022}%
  \BibitemOpen
  \bibfield  {author} {\bibinfo {author} {\bibfnamefont {Ivo~D.}\ \bibnamefont
  {Dinov}}\ and\ \bibinfo {author} {\bibfnamefont {Milen~Velchev}\ \bibnamefont
  {Velev}},\ }\href@noop {} {\emph {\bibinfo {title} {Data Science}}}\
  (\bibinfo  {publisher} {de Gruyter GmbH},\ \bibinfo {address} {Berlin,
  Germany},\ \bibinfo {year} {2022})\BibitemShut {NoStop}%
\bibitem [{\citenamefont {Fredrickson}\ and\ \citenamefont
  {Delaney}(2022)}]{Fredrickson2022}%
  \BibitemOpen
  \bibfield  {author} {\bibinfo {author} {\bibfnamefont {Glenn~H.}\
  \bibnamefont {Fredrickson}}\ and\ \bibinfo {author} {\bibfnamefont {Kris~T.}\
  \bibnamefont {Delaney}},\ }\bibfield  {title} {\enquote {\bibinfo {title}
  {Direct free energy evaluation of classical and quantum many-body systems via
  field-theoretic simulation},}\ }\href@noop {} {\bibfield  {journal} {\bibinfo
   {journal} {Proceedings of the National Academy of Sciences}\ }\textbf
  {\bibinfo {volume} {119}},\ \bibinfo {pages} {e2201804119} (\bibinfo {year}
  {2022})}\BibitemShut {NoStop}%
\bibitem [{\citenamefont {Delaney}\ \emph {et~al.}(2020)\citenamefont
  {Delaney}, \citenamefont {Orland},\ and\ \citenamefont
  {Fredrickson}}]{Fredrickson2020}%
  \BibitemOpen
  \bibfield  {author} {\bibinfo {author} {\bibfnamefont {Kris~T.}\ \bibnamefont
  {Delaney}}, \bibinfo {author} {\bibfnamefont {Henri}\ \bibnamefont {Orland}},
  \ and\ \bibinfo {author} {\bibfnamefont {Glenn~H.}\ \bibnamefont
  {Fredrickson}},\ }\bibfield  {title} {\enquote {\bibinfo {title} {Numerical
  simulation of finite-temperature field theory for interacting bosons},}\
  }\href@noop {} {\bibfield  {journal} {\bibinfo  {journal} {Physical Review
  Letters}\ }\textbf {\bibinfo {volume} {124}},\ \bibinfo {pages} {070601}
  (\bibinfo {year} {2020})}\BibitemShut {NoStop}%
\bibitem [{\citenamefont {Simmons}\ \emph {et~al.}(2023)\citenamefont
  {Simmons}, \citenamefont {Sajjad}, \citenamefont {Keithley}, \citenamefont
  {Mas}, \citenamefont {Tanlimco}, \citenamefont {Nolasco-Martinez},
  \citenamefont {Bai}, \citenamefont {Fredrickson},\ and\ \citenamefont
  {Weld}}]{Fredrickson2023b}%
  \BibitemOpen
  \bibfield  {author} {\bibinfo {author} {\bibfnamefont {Ethan~Q.}\
  \bibnamefont {Simmons}}, \bibinfo {author} {\bibfnamefont {Roshan}\
  \bibnamefont {Sajjad}}, \bibinfo {author} {\bibfnamefont {Kimberlee}\
  \bibnamefont {Keithley}}, \bibinfo {author} {\bibfnamefont {Hector}\
  \bibnamefont {Mas}}, \bibinfo {author} {\bibfnamefont {Jeremy~L.}\
  \bibnamefont {Tanlimco}}, \bibinfo {author} {\bibfnamefont {Eber}\
  \bibnamefont {Nolasco-Martinez}}, \bibinfo {author} {\bibfnamefont {Yifei}\
  \bibnamefont {Bai}}, \bibinfo {author} {\bibfnamefont {Glenn~H.}\
  \bibnamefont {Fredrickson}}, \ and\ \bibinfo {author} {\bibfnamefont
  {David~M.}\ \bibnamefont {Weld}},\ }\bibfield  {title} {\enquote {\bibinfo
  {title} {Thermodynamic engine with a quantum degenerate working fluid},}\
  }\href@noop {} {\bibfield  {journal} {\bibinfo  {journal} {Physical Review
  Research}\ }\textbf {\bibinfo {volume} {5}},\ \bibinfo {pages} {L042009}
  (\bibinfo {year} {2023})}\BibitemShut {NoStop}%
\bibitem [{\citenamefont {Mc{G}arrigle}\ \emph {et~al.}(2023)\citenamefont
  {Mc{G}arrigle}, \citenamefont {Delaney}, \citenamefont {Balents},\ and\
  \citenamefont {Fredrickson}}]{Fredrickson2023c}%
  \BibitemOpen
  \bibfield  {author} {\bibinfo {author} {\bibfnamefont {Ethan~C.}\
  \bibnamefont {Mc{G}arrigle}}, \bibinfo {author} {\bibfnamefont {Kris~T.}\
  \bibnamefont {Delaney}}, \bibinfo {author} {\bibfnamefont {Leon}\
  \bibnamefont {Balents}}, \ and\ \bibinfo {author} {\bibfnamefont {Glenn~H.}\
  \bibnamefont {Fredrickson}},\ }\bibfield  {title} {\enquote {\bibinfo {title}
  {Emergence of a spin microemulsion in spin-orbit coupled {B}ose-{E}instein
  condensates},}\ }\href@noop {} {\bibfield  {journal} {\bibinfo  {journal}
  {Physical Review Letters}\ }\textbf {\bibinfo {volume} {131}},\ \bibinfo
  {pages} {173403} (\bibinfo {year} {2023})}\BibitemShut {NoStop}%
\bibitem [{\citenamefont {Fredrickson}\ and\ \citenamefont
  {Delaney}(2023)}]{Fredrickson2023}%
  \BibitemOpen
  \bibfield  {author} {\bibinfo {author} {\bibfnamefont {Glenn~H.}\
  \bibnamefont {Fredrickson}}\ and\ \bibinfo {author} {\bibfnamefont {Kris~T.}\
  \bibnamefont {Delaney}},\ }\href@noop {} {\emph {\bibinfo {title}
  {Field-Theoretic Simulations in Soft Matter and Quantum Fluids}}}\ (\bibinfo
  {publisher} {Oxford University Press},\ \bibinfo {address} {Oxford, UK},\
  \bibinfo {year} {2023})\BibitemShut {NoStop}%
\bibitem [{\citenamefont {Mullin}(2000)}]{Mullin2000}%
  \BibitemOpen
  \bibfield  {author} {\bibinfo {author} {\bibfnamefont {William~J.}\
  \bibnamefont {Mullin}},\ }\bibfield  {title} {\enquote {\bibinfo {title}
  {Permutation cycles in the {B}ose-{E}instein condensation of a trapped ideal
  gas},}\ }\href@noop {} {\bibfield  {journal} {\bibinfo  {journal} {Physica
  B}\ }\textbf {\bibinfo {volume} {284-288}},\ \bibinfo {pages} {7--8}
  (\bibinfo {year} {2000})}\BibitemShut {NoStop}%
\bibitem [{\citenamefont {Mullin}\ and\ \citenamefont
  {Fern\'{a}ndez}(2003)}]{Mullin2003}%
  \BibitemOpen
  \bibfield  {author} {\bibinfo {author} {\bibfnamefont {William~J.}\
  \bibnamefont {Mullin}}\ and\ \bibinfo {author} {\bibfnamefont {J.~P.}\
  \bibnamefont {Fern\'{a}ndez}},\ }\bibfield  {title} {\enquote {\bibinfo
  {title} {{B}ose–{E}instein condensation, fluctuations, and recurrence
  relations in statistical mechanics},}\ }\href@noop {} {\bibfield  {journal}
  {\bibinfo  {journal} {American Journal of Physics}\ }\textbf {\bibinfo
  {volume} {71}},\ \bibinfo {pages} {661--669} (\bibinfo {year}
  {2003})}\BibitemShut {NoStop}%
\bibitem [{\citenamefont {Vakarchuk}\ and\ \citenamefont
  {Rovenchak}(2001)}]{Vakarchuk2001}%
  \BibitemOpen
  \bibfield  {author} {\bibinfo {author} {\bibfnamefont {I.~O.}\ \bibnamefont
  {Vakarchuk}}\ and\ \bibinfo {author} {\bibfnamefont {A.~A.}\ \bibnamefont
  {Rovenchak}},\ }\bibfield  {title} {\enquote {\bibinfo {title}
  {Thermodynamics of the {B}ose-system with a small number of particles},}\
  }\href@noop {} {\bibfield  {journal} {\bibinfo  {journal} {Condensed Matter
  Physics}\ }\textbf {\bibinfo {volume} {4}},\ \bibinfo {pages} {431--447}
  (\bibinfo {year} {2001})}\BibitemShut {NoStop}%
\bibitem [{\citenamefont {Matsubara}(1951)}]{Matsubara1951}%
  \BibitemOpen
  \bibfield  {author} {\bibinfo {author} {\bibfnamefont {Takeo}\ \bibnamefont
  {Matsubara}},\ }\bibfield  {title} {\enquote {\bibinfo {title}
  {Quantum-statistical theory of liquid helium},}\ }\href@noop {} {\bibfield
  {journal} {\bibinfo  {journal} {Progress of Theoretical Physics}\ }\textbf
  {\bibinfo {volume} {6}},\ \bibinfo {pages} {714--730} (\bibinfo {year}
  {1951})}\BibitemShut {NoStop}%
\bibitem [{\citenamefont {Feynman}(1972)}]{Feynman1972}%
  \BibitemOpen
  \bibfield  {author} {\bibinfo {author} {\bibfnamefont {Richard~P.}\
  \bibnamefont {Feynman}},\ }\href@noop {} {\emph {\bibinfo {title}
  {Statistical Mechanics}}}\ (\bibinfo  {publisher} {W. A. Benjamin},\ \bibinfo
  {address} {New York, USA},\ \bibinfo {year} {1972})\BibitemShut {NoStop}%
\bibitem [{\citenamefont {Barghathi}\ \emph {et~al.}(2020)\citenamefont
  {Barghathi}, \citenamefont {Yu},\ and\ \citenamefont
  {Maestro}}]{DelMaestro2020}%
  \BibitemOpen
  \bibfield  {author} {\bibinfo {author} {\bibfnamefont {Hatem}\ \bibnamefont
  {Barghathi}}, \bibinfo {author} {\bibfnamefont {Jiangyong}\ \bibnamefont
  {Yu}}, \ and\ \bibinfo {author} {\bibfnamefont {Adrian~Del}\ \bibnamefont
  {Maestro}},\ }\bibfield  {title} {\enquote {\bibinfo {title} {Theory of
  non-interacting fermions and bosons in the canonical ensemble},}\ }\href@noop
  {} {\bibfield  {journal} {\bibinfo  {journal} {Physical Review Research}\
  }\textbf {\bibinfo {volume} {2}},\ \bibinfo {pages} {043206} (\bibinfo {year}
  {2020})}\BibitemShut {NoStop}%
\bibitem [{\citenamefont {Kocharovsky}\ and\ \citenamefont
  {Kocharovsky}(2010)}]{Kocharovsky2010}%
  \BibitemOpen
  \bibfield  {author} {\bibinfo {author} {\bibfnamefont {Vitaly~V.}\
  \bibnamefont {Kocharovsky}}\ and\ \bibinfo {author} {\bibfnamefont
  {Vladimir~V.}\ \bibnamefont {Kocharovsky}},\ }\bibfield  {title} {\enquote
  {\bibinfo {title} {Analytical theory of mesoscopic {B}ose-{E}instein
  condensation in an ideal gas},}\ }\href@noop {} {\bibfield  {journal}
  {\bibinfo  {journal} {Physical Review A}\ }\textbf {\bibinfo {volume} {81}},\
  \bibinfo {pages} {033615} (\bibinfo {year} {2010})}\BibitemShut {NoStop}%
\bibitem [{\citenamefont {Tarasov}\ \emph {et~al.}(2014)\citenamefont
  {Tarasov}, \citenamefont {Kocharovsky},\ and\ \citenamefont
  {Kocharovsky}}]{Kocharovsky2014}%
  \BibitemOpen
  \bibfield  {author} {\bibinfo {author} {\bibfnamefont {S.~V.}\ \bibnamefont
  {Tarasov}}, \bibinfo {author} {\bibfnamefont {{Vl}.~V.}\ \bibnamefont
  {Kocharovsky}}, \ and\ \bibinfo {author} {\bibfnamefont {V.~V.}\ \bibnamefont
  {Kocharovsky}},\ }\bibfield  {title} {\enquote {\bibinfo {title} {Universal
  scaling in the statistics and thermodynamics of a {B}ose-{E}instein
  condensation of an ideal gas in an arbitrary trap},}\ }\href@noop {}
  {\bibfield  {journal} {\bibinfo  {journal} {Physical Review A}\ }\textbf
  {\bibinfo {volume} {90}},\ \bibinfo {pages} {033605} (\bibinfo {year}
  {2014})}\BibitemShut {NoStop}%
\bibitem [{\citenamefont {Tarasov}\ \emph {et~al.}(2015)\citenamefont
  {Tarasov}, \citenamefont {Kocharovsky},\ and\ \citenamefont
  {Kocharovsky}}]{Kocharovsky2015}%
  \BibitemOpen
  \bibfield  {author} {\bibinfo {author} {\bibfnamefont {S.~V.}\ \bibnamefont
  {Tarasov}}, \bibinfo {author} {\bibfnamefont {{Vl}.~V.}\ \bibnamefont
  {Kocharovsky}}, \ and\ \bibinfo {author} {\bibfnamefont {V.~V.}\ \bibnamefont
  {Kocharovsky}},\ }\bibfield  {title} {\enquote {\bibinfo {title} {Grand
  canonical versus canonical ensemble: Universal structure of statistics and
  thermodynamics in a critical region of {B}ose–{E}instein condensation of an
  ideal gas in arbitrary trap},}\ }\href@noop {} {\bibfield  {journal}
  {\bibinfo  {journal} {Journal of Statistical Physics}\ }\textbf {\bibinfo
  {volume} {161}},\ \bibinfo {pages} {942--964} (\bibinfo {year}
  {2015})}\BibitemShut {NoStop}%
\bibitem [{\citenamefont {Simon}(2002)}]{Simon2002}%
  \BibitemOpen
  \bibfield  {author} {\bibinfo {author} {\bibfnamefont {Christoph}\
  \bibnamefont {Simon}},\ }\bibfield  {title} {\enquote {\bibinfo {title}
  {Natural entanglement in {B}ose-{E}instein condensates},}\ }\href@noop {}
  {\bibfield  {journal} {\bibinfo  {journal} {Physical Review A}\ }\textbf
  {\bibinfo {volume} {66}},\ \bibinfo {pages} {052323} (\bibinfo {year}
  {2002})}\BibitemShut {NoStop}%
\bibitem [{\citenamefont {Heaney}\ and\ \citenamefont
  {Vedral}(2009)}]{Heaney2009}%
  \BibitemOpen
  \bibfield  {author} {\bibinfo {author} {\bibfnamefont {Libby}\ \bibnamefont
  {Heaney}}\ and\ \bibinfo {author} {\bibfnamefont {Vlatko}\ \bibnamefont
  {Vedral}},\ }\bibfield  {title} {\enquote {\bibinfo {title} {Natural mode
  entanglement as a resource for quantum communication},}\ }\href@noop {}
  {\bibfield  {journal} {\bibinfo  {journal} {Physical Review Letters}\
  }\textbf {\bibinfo {volume} {103}},\ \bibinfo {pages} {200502} (\bibinfo
  {year} {2009})}\BibitemShut {NoStop}%
\bibitem [{\citenamefont {Laflorencie}(2016)}]{Laflorencie2016}%
  \BibitemOpen
  \bibfield  {author} {\bibinfo {author} {\bibfnamefont {Nicolas}\ \bibnamefont
  {Laflorencie}},\ }\bibfield  {title} {\enquote {\bibinfo {title} {Quantum
  entanglement in condensed matter systems},}\ }\href@noop {} {\bibfield
  {journal} {\bibinfo  {journal} {Physics Reports}\ }\textbf {\bibinfo {volume}
  {646}},\ \bibinfo {pages} {1--59} (\bibinfo {year} {2016})}\BibitemShut
  {NoStop}%
\bibitem [{\citenamefont {Schmied}\ \emph {et~al.}(2016)\citenamefont
  {Schmied}, \citenamefont {Bancal}, \citenamefont {Allard}, \citenamefont
  {Fadel}, \citenamefont {Scarani}, \citenamefont {Treutlein},\ and\
  \citenamefont {Sangouard}}]{Schmied2016}%
  \BibitemOpen
  \bibfield  {author} {\bibinfo {author} {\bibfnamefont {Roman}\ \bibnamefont
  {Schmied}}, \bibinfo {author} {\bibfnamefont {Jean-Daniel}\ \bibnamefont
  {Bancal}}, \bibinfo {author} {\bibfnamefont {Baptiste}\ \bibnamefont
  {Allard}}, \bibinfo {author} {\bibfnamefont {Matteo}\ \bibnamefont {Fadel}},
  \bibinfo {author} {\bibfnamefont {Valerio}\ \bibnamefont {Scarani}}, \bibinfo
  {author} {\bibfnamefont {Philipp}\ \bibnamefont {Treutlein}}, \ and\ \bibinfo
  {author} {\bibfnamefont {Nicolas}\ \bibnamefont {Sangouard}},\ }\bibfield
  {title} {\enquote {\bibinfo {title} {Bell correlations in a {B}ose-{E}instein
  condensate},}\ }\href@noop {} {\bibfield  {journal} {\bibinfo  {journal}
  {Science}\ }\textbf {\bibinfo {volume} {352}},\ \bibinfo {pages} {441--444}
  (\bibinfo {year} {2016})}\BibitemShut {NoStop}%
\bibitem [{\citenamefont {Matsen}(2006)}]{Matsen2006}%
  \BibitemOpen
  \bibfield  {author} {\bibinfo {author} {\bibfnamefont {Mark~W.}\ \bibnamefont
  {Matsen}},\ }\bibfield  {title} {\enquote {\bibinfo {title} {Self-consistent
  field theory and its applications},}\ }in\ \href@noop {} {\emph {\bibinfo
  {booktitle} {Soft Matter, Volume 1: Polymer Melts and Mixtures}}},\ \bibinfo
  {editor} {edited by\ \bibinfo {editor} {\bibfnamefont {G.}~\bibnamefont
  {Gompper}}\ and\ \bibinfo {editor} {\bibfnamefont {M.}~\bibnamefont
  {Schick}}}\ (\bibinfo  {publisher} {Wiley-VCH},\ \bibinfo {address}
  {Weinheim},\ \bibinfo {year} {2006})\ pp.\ \bibinfo {pages}
  {87--178}\BibitemShut {NoStop}%
\bibitem [{\citenamefont {Fredrickson}(2006)}]{Fredrickson2006}%
  \BibitemOpen
  \bibfield  {author} {\bibinfo {author} {\bibfnamefont {Glenn~H.}\
  \bibnamefont {Fredrickson}},\ }\href@noop {} {\emph {\bibinfo {title} {The
  Equilibrium Theory of Inhomogeneous Polymers}}}\ (\bibinfo  {publisher}
  {Oxford University Press},\ \bibinfo {address} {New York, NY},\ \bibinfo
  {year} {2006})\BibitemShut {NoStop}%
\bibitem [{\citenamefont {Matsen}(2002)}]{Matsen2002}%
  \BibitemOpen
  \bibfield  {author} {\bibinfo {author} {\bibfnamefont {Mark~W.}\ \bibnamefont
  {Matsen}},\ }\bibfield  {title} {\enquote {\bibinfo {title} {The standard
  {G}aussian model for block copolymer melts},}\ }\href@noop {} {\bibfield
  {journal} {\bibinfo  {journal} {Journal of Physics: Condensed Matter}\
  }\textbf {\bibinfo {volume} {14}},\ \bibinfo {pages} {R21--R47} (\bibinfo
  {year} {2002})}\BibitemShut {NoStop}%
\bibitem [{\citenamefont {Yang}\ \emph {et~al.}(2006)\citenamefont {Yang},
  \citenamefont {Qiu}, \citenamefont {Tang},\ and\ \citenamefont
  {Zhang}}]{Qiu2006}%
  \BibitemOpen
  \bibfield  {author} {\bibinfo {author} {\bibfnamefont {Yuliang}\ \bibnamefont
  {Yang}}, \bibinfo {author} {\bibfnamefont {Feng}\ \bibnamefont {Qiu}},
  \bibinfo {author} {\bibfnamefont {Ping}\ \bibnamefont {Tang}}, \ and\
  \bibinfo {author} {\bibfnamefont {Hongdong}\ \bibnamefont {Zhang}},\
  }\bibfield  {title} {\enquote {\bibinfo {title} {Applications of
  self-consistent field theory in polymer systems},}\ }\href@noop {} {\bibfield
   {journal} {\bibinfo  {journal} {Science in China: Series B Chemistry}\
  }\textbf {\bibinfo {volume} {49}},\ \bibinfo {pages} {21--43} (\bibinfo
  {year} {2006})}\BibitemShut {NoStop}%
\bibitem [{\citenamefont {Schmid}(1998)}]{Schmid1998}%
  \BibitemOpen
  \bibfield  {author} {\bibinfo {author} {\bibfnamefont {Friederike}\
  \bibnamefont {Schmid}},\ }\bibfield  {title} {\enquote {\bibinfo {title}
  {Self-consistent-field theories for complex fluids},}\ }\href@noop {}
  {\bibfield  {journal} {\bibinfo  {journal} {Journal of Physics: Condensed
  Matter}\ }\textbf {\bibinfo {volume} {10}},\ \bibinfo {pages} {8105--8138}
  (\bibinfo {year} {1998})}\BibitemShut {NoStop}%
\bibitem [{\citenamefont {Zhou}\ and\ \citenamefont {Dai}(2018)}]{Zhou2018}%
  \BibitemOpen
  \bibfield  {author} {\bibinfo {author} {\bibfnamefont {Chi-Chun}\
  \bibnamefont {Zhou}}\ and\ \bibinfo {author} {\bibfnamefont {Wu-Sheng}\
  \bibnamefont {Dai}},\ }\bibfield  {title} {\enquote {\bibinfo {title} {A
  statistical mechanical approach to restricted integer partition functions},}\
  }\href@noop {} {\bibfield  {journal} {\bibinfo  {journal} {Journal of
  Statistical Mechanics: Theory and Experiment}\ ,\ \bibinfo {pages} {053111}}
  (\bibinfo {year} {2018})}\BibitemShut {NoStop}%
\bibitem [{\citenamefont {Landsberg}(1961)}]{Landsberg1961}%
  \BibitemOpen
  \bibfield  {author} {\bibinfo {author} {\bibfnamefont {P.~T.}\ \bibnamefont
  {Landsberg}},\ }\href@noop {} {\emph {\bibinfo {title} {Thermodynamics}}}\
  (\bibinfo  {publisher} {Interscience},\ \bibinfo {address} {New York, USA},\
  \bibinfo {year} {1961})\BibitemShut {NoStop}%
\bibitem [{\citenamefont {Park}\ and\ \citenamefont {Kim}(2010)}]{Park2010}%
  \BibitemOpen
  \bibfield  {author} {\bibinfo {author} {\bibfnamefont {Jeong-Hyuck}\
  \bibnamefont {Park}}\ and\ \bibinfo {author} {\bibfnamefont {Sang-Woo}\
  \bibnamefont {Kim}},\ }\bibfield  {title} {\enquote {\bibinfo {title}
  {Thermodynamic instability and first-order phase transition in an ideal
  {B}ose gas},}\ }\href@noop {} {\bibfield  {journal} {\bibinfo  {journal}
  {Physical Review A}\ }\textbf {\bibinfo {volume} {81}},\ \bibinfo {pages}
  {063636} (\bibinfo {year} {2010})}\BibitemShut {NoStop}%
\bibitem [{\citenamefont {Ballentine}(1970)}]{Ballentine1970}%
  \BibitemOpen
  \bibfield  {author} {\bibinfo {author} {\bibfnamefont {Leslie~E.}\
  \bibnamefont {Ballentine}},\ }\bibfield  {title} {\enquote {\bibinfo {title}
  {The statistical interpretation of quantum mechanics},}\ }\href@noop {}
  {\bibfield  {journal} {\bibinfo  {journal} {Rev. Mod. Phys.}\ }\textbf
  {\bibinfo {volume} {42}},\ \bibinfo {pages} {358--381} (\bibinfo {year}
  {1970})}\BibitemShut {NoStop}%
\bibitem [{\citenamefont {Peres}(2002)}]{Peres2002}%
  \BibitemOpen
  \bibfield  {author} {\bibinfo {author} {\bibfnamefont {Asher}\ \bibnamefont
  {Peres}},\ }\href@noop {} {\emph {\bibinfo {title} {Quantum Theory: Concepts
  and Methods}}}\ (\bibinfo  {publisher} {Kluwer},\ \bibinfo {address} {New
  York, USA},\ \bibinfo {year} {2002})\BibitemShut {NoStop}%
\bibitem [{\citenamefont {LaCour}(2009)}]{LaCour2009}%
  \BibitemOpen
  \bibfield  {author} {\bibinfo {author} {\bibfnamefont {Brian~R.}\
  \bibnamefont {LaCour}},\ }\bibfield  {title} {\enquote {\bibinfo {title}
  {Quantum contextuality in the {M}ermin-{P}eres square: {A} hidden-variable
  perspective},}\ }\href@noop {} {\bibfield  {journal} {\bibinfo  {journal}
  {Physical Review A}\ }\textbf {\bibinfo {volume} {79}},\ \bibinfo {pages}
  {012102} (\bibinfo {year} {2009})}\BibitemShut {NoStop}%
\bibitem [{\citenamefont {Bell}(1964)}]{Bell1964}%
  \BibitemOpen
  \bibfield  {author} {\bibinfo {author} {\bibfnamefont {John~S.}\ \bibnamefont
  {Bell}},\ }\bibfield  {title} {\enquote {\bibinfo {title} {On the {Einstein}
  {Podolsky} {Rosen} paradox},}\ }\href@noop {} {\bibfield  {journal} {\bibinfo
   {journal} {Physics}\ }\textbf {\bibinfo {volume} {1}},\ \bibinfo {pages}
  {195--200} (\bibinfo {year} {1964})}\BibitemShut {NoStop}%
\bibitem [{\citenamefont {Myrvold}(2016)}]{Myrvold2016}%
  \BibitemOpen
  \bibfield  {author} {\bibinfo {author} {\bibfnamefont {Wayne~C.}\
  \bibnamefont {Myrvold}},\ }\bibfield  {title} {\enquote {\bibinfo {title}
  {Lessons of {B}ell’s {T}heorem: {N}onlocality, yes; {A}ction at a distance,
  not necessarily},}\ }in\ \href@noop {} {\emph {\bibinfo {booktitle} {Quantum
  {N}onlocality and {R}eality: 50 {Y}ears of {B}ell's {T}heorem}}},\ \bibinfo
  {editor} {edited by\ \bibinfo {editor} {\bibfnamefont {Mary}\ \bibnamefont
  {Bell}}\ and\ \bibinfo {editor} {\bibfnamefont {Shan}\ \bibnamefont {Gao}}}\
  (\bibinfo  {publisher} {Cambridge {U}niversity {P}ress},\ \bibinfo {address}
  {Cambridge, UK},\ \bibinfo {year} {2016})\ pp.\ \bibinfo {pages}
  {238--260}\BibitemShut {NoStop}%
\bibitem [{\citenamefont {Wharton}(2016)}]{Wharton2016}%
  \BibitemOpen
  \bibfield  {author} {\bibinfo {author} {\bibfnamefont {Ken}\ \bibnamefont
  {Wharton}},\ }\bibfield  {title} {\enquote {\bibinfo {title} {Towards a
  realistic parsing of the {F}eynman path integral},}\ }\href@noop {}
  {\bibfield  {journal} {\bibinfo  {journal} {Quanta}\ }\textbf {\bibinfo
  {volume} {5}},\ \bibinfo {pages} {1--11} (\bibinfo {year}
  {2016})}\BibitemShut {NoStop}%
\bibitem [{\citenamefont {Tyagi}\ and\ \citenamefont
  {Wharton}(2022)}]{Wharton2022}%
  \BibitemOpen
  \bibfield  {author} {\bibinfo {author} {\bibfnamefont {Narayani}\
  \bibnamefont {Tyagi}}\ and\ \bibinfo {author} {\bibfnamefont {Ken}\
  \bibnamefont {Wharton}},\ }\bibfield  {title} {\enquote {\bibinfo {title}
  {Spacetime path integrals for entangled states},}\ }\href@noop {} {\bibfield
  {journal} {\bibinfo  {journal} {Foundations of Physics}\ }\textbf {\bibinfo
  {volume} {52}},\ \bibinfo {pages} {9} (\bibinfo {year} {2022})}\BibitemShut
  {NoStop}%
\bibitem [{\citenamefont {Wharton}\ and\ \citenamefont
  {Liu}(2023)}]{Wharton2023}%
  \BibitemOpen
  \bibfield  {author} {\bibinfo {author} {\bibfnamefont {Ken}\ \bibnamefont
  {Wharton}}\ and\ \bibinfo {author} {\bibfnamefont {Raylor}\ \bibnamefont
  {Liu}},\ }\bibfield  {title} {\enquote {\bibinfo {title} {Entanglement and
  the path integral},}\ }\href@noop {} {\bibfield  {journal} {\bibinfo
  {journal} {Foundations of Physics}\ }\textbf {\bibinfo {volume} {53}},\
  \bibinfo {pages} {23} (\bibinfo {year} {2023})}\BibitemShut {NoStop}%
\bibitem [{\citenamefont {Eisert}(2010)}]{Eisert2010}%
  \BibitemOpen
  \bibfield  {author} {\bibinfo {author} {\bibfnamefont {J.}~\bibnamefont
  {Eisert}},\ }\bibfield  {title} {\enquote {\bibinfo {title} {Area laws for
  the entanglement entropy},}\ }\href@noop {} {\bibfield  {journal} {\bibinfo
  {journal} {Reviews of Modern Physics}\ }\textbf {\bibinfo {volume} {82}},\
  \bibinfo {pages} {277--306} (\bibinfo {year} {2010})}\BibitemShut {NoStop}%
\bibitem [{\citenamefont {Alba}\ \emph {et~al.}(2013)\citenamefont {Alba},
  \citenamefont {Haque},\ and\ \citenamefont {L\"{a}uchli}}]{Alba2013}%
  \BibitemOpen
  \bibfield  {author} {\bibinfo {author} {\bibfnamefont {Vincenzo}\
  \bibnamefont {Alba}}, \bibinfo {author} {\bibfnamefont {Masudul}\
  \bibnamefont {Haque}}, \ and\ \bibinfo {author} {\bibfnamefont {Andreas~M.}\
  \bibnamefont {L\"{a}uchli}},\ }\bibfield  {title} {\enquote {\bibinfo {title}
  {Entanglement spectrum of the two-dimensional {B}ose-{H}ubbard model},}\
  }\href@noop {} {\bibfield  {journal} {\bibinfo  {journal} {Physical Review
  Letters}\ }\textbf {\bibinfo {volume} {110}},\ \bibinfo {pages} {260403}
  (\bibinfo {year} {2013})}\BibitemShut {NoStop}%
\bibitem [{\citenamefont {Ding}\ and\ \citenamefont {Yang}(2009)}]{Ding2009}%
  \BibitemOpen
  \bibfield  {author} {\bibinfo {author} {\bibfnamefont {Wenxin}\ \bibnamefont
  {Ding}}\ and\ \bibinfo {author} {\bibfnamefont {Kun}\ \bibnamefont {Yang}},\
  }\bibfield  {title} {\enquote {\bibinfo {title} {Entanglement entropy and
  mutual information in {B}ose-{E}instein condensates},}\ }\href@noop {}
  {\bibfield  {journal} {\bibinfo  {journal} {Physical Review A}\ }\textbf
  {\bibinfo {volume} {80}},\ \bibinfo {pages} {012329} (\bibinfo {year}
  {2009})}\BibitemShut {NoStop}%
\bibitem [{\citenamefont {Buonacorsi}\ \emph {et~al.}(2019)\citenamefont
  {Buonacorsi}, \citenamefont {Cai}, \citenamefont {Ramirez}, \citenamefont
  {Willick}, \citenamefont {Walker}, \citenamefont {Li}, \citenamefont {Shaw},
  \citenamefont {Xu}, \citenamefont {Benjamin},\ and\ \citenamefont
  {Baugh}}]{Baugh2019}%
  \BibitemOpen
  \bibfield  {author} {\bibinfo {author} {\bibfnamefont {Brandon}\ \bibnamefont
  {Buonacorsi}}, \bibinfo {author} {\bibfnamefont {Zhenyu}\ \bibnamefont
  {Cai}}, \bibinfo {author} {\bibfnamefont {Eduardo~B.}\ \bibnamefont
  {Ramirez}}, \bibinfo {author} {\bibfnamefont {Kyle~S.}\ \bibnamefont
  {Willick}}, \bibinfo {author} {\bibfnamefont {Sean~M.}\ \bibnamefont
  {Walker}}, \bibinfo {author} {\bibfnamefont {Jiahao}\ \bibnamefont {Li}},
  \bibinfo {author} {\bibfnamefont {Benjamin~D.}\ \bibnamefont {Shaw}},
  \bibinfo {author} {\bibfnamefont {Xiaosi}\ \bibnamefont {Xu}}, \bibinfo
  {author} {\bibfnamefont {Simon~C.}\ \bibnamefont {Benjamin}}, \ and\ \bibinfo
  {author} {\bibfnamefont {Jonathan}\ \bibnamefont {Baugh}},\ }\bibfield
  {title} {\enquote {\bibinfo {title} {Network architecture for a topological
  quantum computer in silicon},}\ }\href@noop {} {\bibfield  {journal}
  {\bibinfo  {journal} {Quantum Science and Technology}\ }\textbf {\bibinfo
  {volume} {4}},\ \bibinfo {pages} {025003} (\bibinfo {year}
  {2019})}\BibitemShut {NoStop}%
\end{thebibliography}%

\end{document}